\documentclass[aps,prl,superscriptaddress,reprint,floatfix]{revtex4-2}

\usepackage{dcolumn}
\usepackage{bm}
\usepackage{float}
\usepackage{braket}
\usepackage{svg}
\usepackage{amsmath}
\usepackage{amssymb} 
\usepackage{graphicx}
\usepackage{caption}
\usepackage[font=small, justification=Justified]{caption}
\usepackage{subcaption}
\usepackage[export]{adjustbox}
\usepackage{hyperref}
\hypersetup{
    colorlinks=true,
    linkcolor=blue,
    citecolor=blue,
    urlcolor=blue,
    }
\usepackage[dvipsnames]{xcolor}
\usepackage{xspace}
\usepackage{textcomp} 

\begin{document}

\preprint{APS/123-QED}

\title{Quantum Nonlinear Response of Emitter Lattices}

\author{Blas Durá-Azorín}
\affiliation{Departamento de Física Teórica de la Materia Condensada, Universidad Autónoma de Madrid, E-28049 Madrid, Spain}
\affiliation{Condensed Matter Physics Center (IFIMAC), Universidad Autónoma de Madrid, E-28049 Madrid, Spain.} 
\affiliation{Instituto de Qu\'imica F\'isica Blas Cabrera (IQF), CSIC, E-28006 Madrid, Spain}
\author{Antonio I. Fernández-Domínguez}
\email{a.fernandez-dominguez@uam.es}
\affiliation{Departamento de Física Teórica de la Materia Condensada, Universidad Autónoma de Madrid, E-28049 Madrid, Spain}
\affiliation{Condensed Matter Physics Center (IFIMAC), Universidad Autónoma de Madrid, E-28049 Madrid, Spain.}
\author{Alejandro Manjavacas}
\email{a.manjavacas@csic.es}
\affiliation{Instituto de Qu\'imica F\'isica Blas Cabrera (IQF), CSIC, E-28006 Madrid, Spain}

\begin{abstract}
We theoretically investigate the emergence of quantum nonlinearities in the optical response of lattices of two-level quantum emitters coherently driven by a laser. For subwavelength lattice periods, where the system behaves as a quantum metasurface, we find that a resonant incident plane wave can populate excitonic Bloch states with parallel wavevectors different from the incident field, including those lying outside the light cone. Closely related to resonance fluorescence, the far-field emission from the system in the strong-driving regime is dominated by a broadband background of photons spanning a wide range of frequencies and wavevectors. Moreover, we show that, for periods approaching the driving wavelength, the emitter lattice enters in a bistable regime due to the renormalization of the driving rate, in striking contrast with its classical (bosonic) analog.  This bistable behavior enables the selective activation and deactivation of the optical quantum nonlinearities of the system.
\end{abstract}

\maketitle

For approximately the past two decades \cite{GarciadeAbjo07_RMP}, periodic lattices of metallic nanostructures have been the subject of an intense theoretical and experimental attention due to their ability to support lattice resonances \cite{Kravets18_CR,Wang18_MT}, which give rise to extraordinary field enhancements and quality factors, much larger than those of the individual nanostructures \cite{Manjavacas19_ACSNano, LeVan19_AOM, Cuartero20_ACSNano, BinAlam21_NC}. More recently, advances in the creation and manipulation of quantum emitter (QE) ensembles \cite{Barredo16_Science,Srakaew23_NatPhys, Rui20_Nat, OhldeMello19_PRL} have sparked intense research into exploring these collective modes in the quantum realm. Many studies have focused on the single excitation limit, which allows for a linearization of the emitters. Within this approximation, it has been shown that QE lattices and metasurfaces can implement strong magnetic responses \cite{Alaee20_PRL}, toroidal and anapole excitations \cite{Ballantine20_PRL}, deeply subradiant modes \cite{Asenjo17_PRX} or bound states in the continuum \cite{BlancodePaz23_PRR}. Beyond the single excitation manifold, strong nonlinear phenomena ranging from bistability phases to anomalous transmission effects \cite{Parmee21_PRA, Ruostekoski23_PRA, Bettles20_ComPhys} have been reported. In parallel, theoretical frameworks have been developed to systematically account for exciton anharmonicity corrections \cite{Pedersen24_PRR}. More recently, the quantum-optical potential of polaritonic crystals coupling QE and optical lattices, have been also shown \cite{Lindel25_ARXIV}. For deeply subwavelength QE metasurfaces, quantum correlations and many-body effects due to large dipole-dipole interactions have been predicted \cite{Scarlatella24_PRA}. All these features convert QE lattices in promising candidates for applications in quantum information technologies \cite{Bekenstein20_NatPhys, SantiagoCruz22_Science}, single photon storage \cite{Asenjo17_PRX, Ballantine21_PRX}, and quantum metrology and sensing \cite{Qu19_PRA,Ostermann13_PRL}.

In this Letter, we explore the impact of quantum nonlinearities in the optical response of coherently driven QE lattices. We find that, in contrast to the classical (bosonic) case, the incident plane wave populates not only the excitonic Bloch states (BSs) with the same parallel wavevector as the driving laser, but a continuum of states with different parallel wavevectors, $\mathbf{k}_\parallel$, including those lying outside the light cone. This phenomenon arises purely from the exciton anharmonicity of the QEs, which is also responsible for effects akin to resonance fluorescence in single QEs \cite{Mollow69_PR, Wu75_PRL, Cohen-Tannoudji77_JoPB, Kavokin17_book, delValle10_PRL}. We also demonstrate the emission of an incoherent background of photons, covering all wavevectors and frequencies, which, under strong-driving conditions, dominates over the Rayleigh (classical) contribution taking place through radiative diffraction orders. Next, we examine the influence of quantum nonlinearities on the emission spectrum of the lattice. These nonlinearities govern the response of the QEs to the coherent driving, effectively renormalizing the laser amplitude and inducing a bistable phase at lattice periods approaching their natural wavelength \cite{Parmee21_PRA}. We analyze the bistable spectrum emerging in this regime and demonstrate that, through the abrupt suppression and restoration of the effective driving, it is possible to selective activate the nonlinear response of the QE lattice.

\begin{figure}[t!]
\centering
\includegraphics[width=0.9\columnwidth]{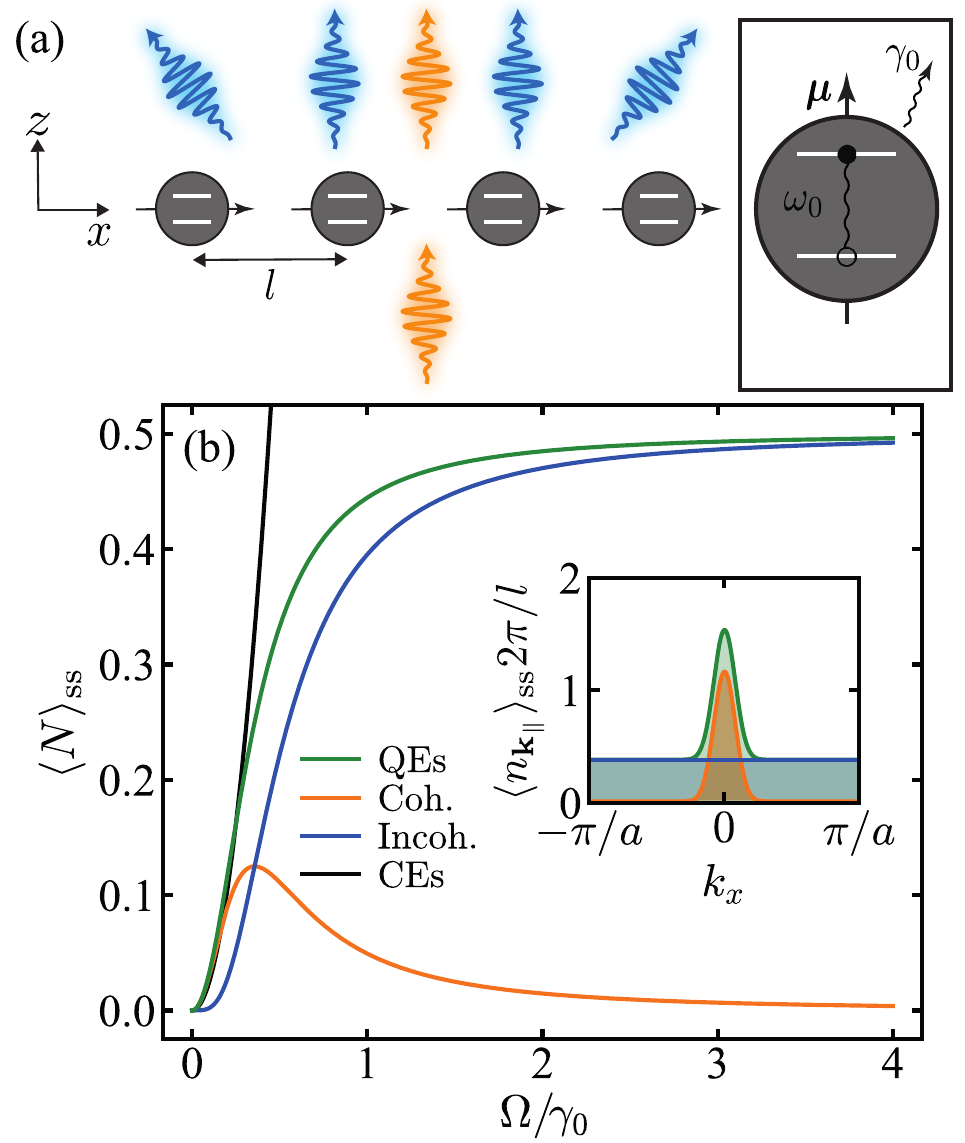}
 \caption{(a) Schematics of a QE lattice with period $l$ lying in the $xy$ plane and excited with an $x$-polarized laser propagating in the $z$ axis. Orange photons indicate emission at $\omega_{\rm L}$ and $\bold{k}_{\rm L,\parallel}$, whereas blue photons highlight emission at other frequencies and wavevectors. The inset shows an schematics of an individual QE. (b) Population per emitter (green), along with its coherent (orange) and incoherent (blue) contributions, as a function of the driving rate for $\Delta=0$. For comparison, the black curve shows $\langle N\rangle_{\rm ss}$ for a lattice of CEs. The inset displays the population distribution within the 1BZ for $k_y=0$, $\Omega=2\gamma_0$, and $\Delta=0$, using the same color scheme.}
\label{fig1}
\end{figure}

The starting point of our theoretical study is the description of the QEs, represented as distinguishable two-level systems characterized by lower quantum operators $\sigma_i$ ($i=1,2,...$) that satisfy a pseudo-spin algebra, $[\sigma_i, \sigma^\dagger_j]=\delta_{ij}(1-2\sigma^\dagger_i \sigma_i)$ and $[\sigma_i, \sigma_j]=0$. We treat them as point-dipoles with dipole moment operator $\boldsymbol{\mu}_i=\boldsymbol{\mu}(\sigma_i + \sigma^\dagger_i)$ and natural frequency $\omega_0$. We neglect dephasing and nonradiative effects, assuming that the emitters present a radiative-limited lifetime, $\gamma^{-1}_0$. The dynamics of the system are given by a Lindblad master equation (see the Appendix for details), $\dot{\rho}={\rm i}[\rho, H/\hbar] + \sum_{i,j}(\gamma_{ij}/2)L_{\sigma_i,\sigma_j}(\rho)$ \cite{Open03_book, Dung02_PRA}. Here, $\rho$ is the density matrix representing the quantum state of the lattice, $H$ is the Hamiltonian
\begin{align}\nonumber
    H/\hbar &=-\Delta \sum_i \sigma_i^\dagger \sigma_i + \sum_{i}\sum_{j\neq i}g_{ij}\sigma_i^\dagger \sigma_j \\
    &+ \sum_i\left[\Omega e^{{\rm i}\bold{k}_{\rm L,\parallel}\cdot \bold{r}_i}\sigma_i^\dagger + \Omega^* e^{-{\rm i}\bold{k}_{\rm L,\parallel}\cdot \bold{r}_i}\sigma_i\right], 
\end{align}
in the rotating frame of the laser and within the rotating wave approximation, and $\Delta=\omega_{\rm L} - \omega_0$ is the detuning between the laser and the natural frequency of the QEs. The parameters $g_{ij}$ and $\gamma_{ij}$ represent the coherent and dissipative coupling strengths, given by the electromagnetic dyadic Green tensor in free space \cite{Dung02_PRA, Novotny12_book}, and $\Omega=(\bf{E}_{\rm L}\cdot\boldsymbol{\mu})/\hbar$ is the driving rate, with $\bf{E}_{\rm L}$ being the amplitude of the laser plane wave and $\bold{k}_{\rm L,\parallel}$ its wavevector. $L_{\sigma_i, \sigma_j}$ are Lindblad superoperators accounting for the QE spontaneous decay and dissipative interactions. From this point onward, all our calculations correspond to the steady-state of the system under external driving, for which $\dot{\rho}=0$.

Under strong coherent driving, the exciton population in a single QE is characterized by its saturation at $1/2$ \cite{Kavokin17_book}. In this regime, the QE power spectrum is composed of two terms: a coherent contribution at the laser frequency, usually known as Rayleigh scattering, and an incoherent one, the so-called Mollow triplet \cite{Mollow69_PR, Kavokin17_book, Casalengua23_PhysScript}. The latter is formed by emission peaks at $\omega_{\rm L}$ and $\sim \omega_{\rm L} \pm 2\Omega$, which originate from two-photon transitions involving virtual states \cite{Cohen-Tannoudji77_JoPB, Aspect80_PRL, Casalengua23_PhysScript}. The optical properties of a QE lattice, as sketched in Fig.~\ref{fig1}(a), can be understood in similar terms. In particular, we consider a square lattice with subwavelength period, $l=0.5\lambda_0$ ($\lambda_0=2\pi c/\omega_0$) placed within the $xy$ plane and excited by a resonant plane wave propagating along the $z$ axis with $\bold{k}_{\rm L, \parallel}=0$ and polarized along $\hat{\boldsymbol{\mu}}=\boldsymbol{\mu}/\mu$ (which we set to be $x$ axis without loss of generality). We focus our attention on the BSs supported by the lattice that are created by the Bloch operators $\sigma^\dagger_{\bold{k}_\parallel}=l/(2\pi) \sum_i e^{{\rm i} \bold{k}_\parallel \cdot \bold{r}_i}\sigma^\dagger_i$, which satisfy $[\sigma^\dagger_{\bold{k}_\parallel},\sigma^\dagger_{\bold{k}^\prime_\parallel}]=0$ and
$[\sigma_{\bold{k}_\parallel},\sigma^\dagger_{\bold{k}^\prime_\parallel}]=\delta(\bold{k}_\parallel - \bold{k}^\prime_\parallel) - 2\mathcal{O}_{\bold{k}_\parallel, \bold{k}^\prime_\parallel}$ with $\mathcal{O}_{\bold{k}_\parallel, \bold{k}^\prime_\parallel}=l^2/(4\pi^2)\sum_{i}e^{-{\rm i}(\bold{k}_\parallel - \bold{k}^\prime_\parallel) \cdot \bold{r}_i}\sigma_i^\dagger\sigma_i$, as described in the Appendix. This expression underscores that BS operators do not obey bosonic or fermionic commutation rules, but different relations that originate from their collective nature and the spin-algebra of the QEs. Importantly, among other properties, the operator above satisfies $\mathcal{O}_{\bold{k}_\parallel, \bold{k}^\prime_\parallel}\ket{\bold{k}^{\prime \prime}_\parallel}=l^2/(4\pi^2)\ket{\bold{k}^{\prime \prime}_\parallel + \bold{k}^{\prime}_\parallel - \bold{k}_\parallel}$, effectively exchanging excitations among different BSs in the lattice. Note that this effect is not due to emitter-emitter interactions, but arises solely from the BS algebra, which reflects the quantum anharmonicity of the QEs.

In order to gain insight into BS excitation by the incoming laser wave, we calculate $\langle n_{\bold{k}_\parallel} \rangle_{\rm ss} = l/(2\pi)\int_{\rm 1BZ} \langle \sigma^\dagger_{\bold{k}_\parallel}\sigma_{\bold{k}^\prime_\parallel} \rangle_{\rm ss} \mathrm{d}\bold{k}^\prime_\parallel$, which describes the population distribution within the first Brillouin zone (1BZ) of the QE lattice. We perform our analysis within the mean-field approximation (MFA) \cite{Parmee21_PRA, Ruostekoski23_PRA}, where we neglect quantum correlations between different emitters, i.e. $\langle \sigma_i^\dagger\sigma_j \rangle \approx \langle\sigma_i^\dagger\rangle\langle \sigma_j \rangle$ for $i\neq j$. This allows us to map the lattice of interacting QEs into a lattice of noninteracting ones, with a renormalized coherent driving, $\Omega_{\rm eff}$, that is a function of $\Omega$,  $\Delta$, and the lattice sum, $\mathcal{G}(\bold{k}_{\rm L, \parallel},\omega_0)$, encoding all the QE interactions \cite{Zundel22_ACSPhot}. We obtain
\begin{align}\nonumber
    \langle n_{\bold{k}_\parallel} \rangle_{\rm ss}&=\frac{4|\Omega_{\rm eff}|^2 (\gamma_0^2+ 4\Delta^2)}{(\gamma_0^2+4\Delta^2+8|\Omega_{\rm eff}|^2)^2} \frac{2\pi}{l}\delta(\bold{k}_\parallel-\bold{k}_{\rm L, \parallel})\\ 
    &+\frac{32|\Omega_{\rm eff}|^4}{(\gamma_0^2+4\Delta^2+8|\Omega_{\rm eff}|^2)^2}\frac{l}{2\pi},
\label{n k}
\end{align}
which reveals two different contributions to the exciton population. The first (coherent) term corresponds to the population of the BS with the incident wavevector, $\bold{k}_{\rm L,\parallel}$. It describes the classical, linear response of the QE lattice. The second term unveils an incoherent mechanism of BS population, which is independent of $\bold{k}_{\rm L, \parallel}$ and yields a nonzero population even for states lying outside the radiative light cone ($|\bold{k}_\parallel| > k=\omega/c$). It originates from the intrinsic quantum nonlinearity of the QEs, for which $\langle \sigma^\dagger_i \sigma_i \rangle_{\rm ss} \neq \langle \sigma^\dagger_i \rangle_{\rm ss} \langle \sigma_i \rangle_{\rm ss} $. In the limit $\Omega_{\rm eff}\rightarrow \infty$, only this second term survives, approaching $1/2$.  This phenomenology closely resembles that of resonance fluorescence, as the non-bosonic character of $\sigma_{\bold{k}_\parallel}$ enables two-photon processes within the QE lattice that transfer the population to wavevector-detuned BSs. These results generalize the ones obtained previously with other methods in the deep subwavelength regime \cite{Scarlatella25_ARXIV}. 

The green curve in Fig.~\ref{fig1}(b) represents the exciton population per emitter, $\langle N \rangle_{\rm ss}=l/(2\pi)\int_{\rm 1BZ}\langle n_{\bold{k_\parallel}} \rangle_{\rm ss} \mathrm{d}\bold{k}_\parallel$, for the QE lattice depicted in panel (a) as a function of $\Omega$ for $\Delta=0$. The orange and blue curves show its coherent and incoherent contributions, respectively, calculated from the first and second terms in Eq.~\eqref{n k}. For comparison, the population per site in its bosonic counterpart, an array of classical emitters (CEs), is shown in black. This population arises solely from a Rayleigh term (see the Appendix) and only BSs with $\bold{k}_\parallel=\bold{k}_{\rm L,\parallel}$ are excited in the system. As a result, $\langle N \rangle_{\rm ss}$ exhibits a quadratic dependence with $\Omega$, reminiscent of a single harmonic oscillator. In contrast, the dominance of the incoherent population mechanism in the QE lattice for moderate drivings (well below the saturation,  $\langle N \rangle_{\rm ss}\rightarrow 1/2$) limits the validity of its bosonic description to the regime $\Omega<<\gamma_0$. For completeness, the inset of Fig.~\ref{fig1}(b) displays $\langle n_{\bold{k}_\parallel} \rangle_{\rm ss}$, obtained from Eq.~\eqref{n k} at $k_y=0$, $\Omega=2\gamma_0$, and $\Delta=0$. To facilitate the visualization of these results, we replace the Dirac delta functions with Gaussian functions of width $2\pi/L$, where $L$ is an effective total length of the lattice,  arbitrarily set to $L=25l$ (mimicking a finite array with $25\times25$ emitters). The green shaded area corresponds to the total population, while orange and blue areas represent the coherent and incoherent contributions, respectively. The former appears as a peak centered at $\bold{k}_{\rm L,\parallel}=0$, while the latter forms a flat continuum in reciprocal space.

\begin{figure}[t!]
\centering
\includegraphics[width=0.9\columnwidth]{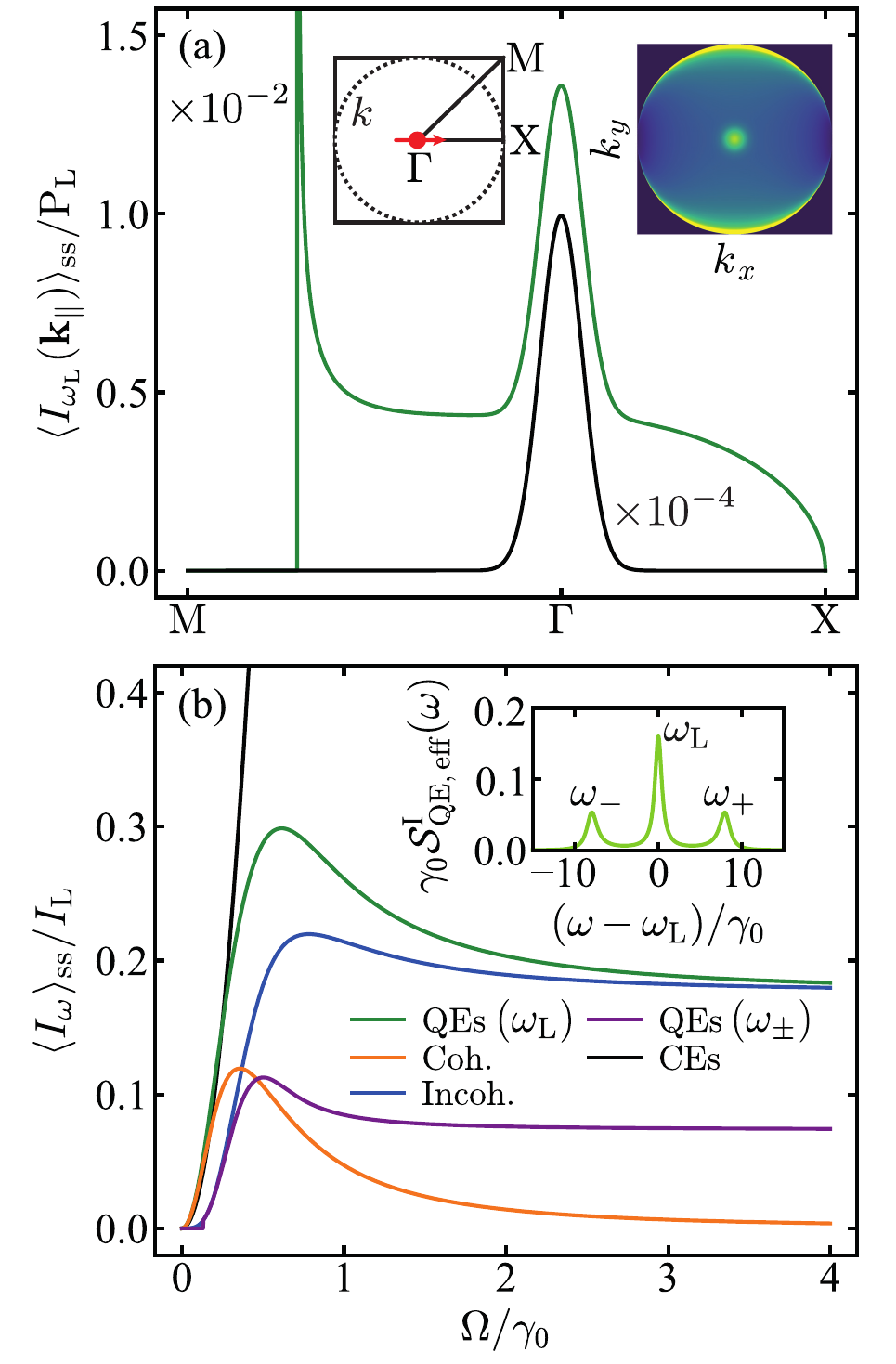}
\caption{(a) Intensity integrated within the central peak of the Mollow-like spectrum, $\langle I_{\omega_{\rm L}} (\bold{k}_\parallel)\rangle_{\rm ss}$, evaluated along the path indicated in the inset for $\Omega=4\gamma_0$ (green). For comparison, the black curve corresponds to a CE lattice. The right inset displays $\langle I_{\omega_{\rm L}} (\bold{k}_\parallel)\rangle_{\rm ss}$ across the entire 1BZ. (b) Total intensity, $\langle I_\omega\rangle_{\rm ss}$, around $\omega_{\rm L}$ (green), and  $\omega_{\pm}$ (purple). The black curve again corresponds to the CE lattice, while the orange and blue curves represent the coherent and incoherent contributions to $\langle I_{\omega_{\rm L}}\rangle_{\rm ss}$, respectively. The inset shows $\mathcal{S}^{\rm I}_{\rm QE, eff}(\omega)$ for $\Omega=4\gamma_0$. In all cases, we assume $\Delta=0$.}
\label{fig2}
\end{figure}

Next, we focus our attention on the photon emission characteristics of the QE lattice. To that end, we calculate the intensity per unit frequency emitted by each BSs that crosses a plane parallel to the lattice (details of the calculation are explained in the Appendix),

\begin{widetext}
\begin{align}
 \langle I(\bold{k}_\parallel,\omega) \rangle_{\rm ss}=\frac{4|\Omega_{\rm eff}|^2(\gamma_0^2+4\Delta^2)}{(\gamma_0^2+4\Delta^2+8|\Omega_{\rm eff}|^2)^2}\sum_{\bold{g}\in \text{rad}}k_{z,\bold{g}} M_{\bold{g}}(\bold{k}_{\parallel},\omega) \delta(\omega-\omega_{\rm L})\delta(\bold{k}_\parallel - \bold{k}_{\rm L, \parallel})+ \frac{l^2}{4\pi^2} k_zM_{\bold 0}(\bold{k}_{\parallel},\omega)\mathcal{S}^{\rm I}_{\rm QE, eff}(\omega).
\label{P k}
\end{align}
\end{widetext}
Here, $k_{z,\bold{g}}=\sqrt{k^2-|\bold{k}_\parallel + \bold{g}|^2}$ and $\bold{g}$ denotes the reciprocal lattice vectors. The function 
$M_{\bold{g}}(\bold{k}_\parallel,\omega)=\left[k^2\boldsymbol{\mu}-\bold{k}_{\bold{g}}(\bold{k}_{\bold{g}}\cdot\boldsymbol{\mu})\right]^2 c^2/(4\omega\varepsilon_0 l^4 k_{z,\bold{g}}^2)$ weights the contribution of the wavevector $\bold{k}_{\bold{g}}=(\bold{k}_\parallel + \bold{g}, k_{z,\bold{g}})$ and frequency $\omega$ to the emitted intensity. The first term in Eq.~\eqref{P k}, in which the sum runs over the radiative reciprocal lattice vectors (i.e. those for which $|\bold{k}_\parallel + \bold{g}|\leq k$), corresponds to the coherent emission associated with the diffraction of the incident laser photons. The second, incoherent term is again independent of $\bold{k}_{\rm L, \parallel}$ and includes the Mollow spectrum of a single QE, effectively renormalized by the MFA, 
\begin{align}
\mathcal{S}^{\rm I}_{\rm QE, eff}(\omega)=\sum_{p=1}^3\frac{L_{\mathrm{eff},p}}{\pi}\frac{\gamma_{\mathrm{eff},p}/2}{(\omega-\omega_{\rm L}-\omega_{\mathrm{eff},p})^2 + \left(\frac{\gamma_{\mathrm{eff},p}}{2}\right)^2}.
\label{Si}
\end{align}
Note that we have neglected the dispersive contribution to $ \mathcal{S}^{\rm I}_{\rm QE, eff}(\omega)$ \cite{delValle2009}, which is negligible in our calculations, as can be inferred from the inset of Fig.~\ref{fig2}(b). The parameters $\omega_{\mathrm{eff},p}$, $\gamma_{\mathrm{eff},p}$, and $L_{\mathrm{eff},p}$ denote the central frequencies, linewidths, and weighting factors of the Lorentzians comprising the spectrum. Thus, the second term in Eq.~\eqref{P k} reveals the presence of a background of photon emission occurring over a broad range of $\bold{k}_\parallel$ and $\omega$, resulting from the incoherent population of BSs in the QE lattice.

The green curve in Fig.~\ref{fig2}(a) shows $\langle I_{\omega_{\rm L}} (\bold{k}_\parallel)\rangle_{\rm ss} =\int_{\omega_{\rm L}}\langle I(\bold{k}_\parallel,\omega) \rangle_{\rm ss}\mathrm{d}\omega$, which represents the intensity emitted by the QE lattice, integrated over the central peak of the Mollow-like spectrum (i.e., in the frequency window between $\omega_{\rm L} - \gamma_0$ and $\omega_{\rm L} + \gamma_0$). The calculation corresponds to the same QE lattice as Fig.~\ref{fig1}, for $\Omega=4\gamma_0$ and $\Delta=0$. As before, we replace the Dirac delta functions in $\bold{k}_\parallel$ with Gaussian functions of width $2\pi/L$. We normalize $\langle I_{\omega_{\rm L}}(\bold{k}_\parallel) \rangle_{\rm ss}$ using $P_{\rm L}=\omega_{\rm L}^4 \mu^2/(12\pi\varepsilon_0c^3)$, which corresponds to the emission by a single QE. We observe that, apart from the peak at the $\Gamma$-point corresponding to zeroth-order diffraction, substantial emission occurs across a continuum of $\bold{k}_\parallel$ within the light cone. 
For comparison, we include the intensity emitted by the corresponding CE lattice, plotted as a black curve. In this case, the emission is restricted to the zeroth-order diffraction peak, which exhibits a finite linewidth due to the finite-size effects encoded in $L$. The wavevector-detuned emission from the QE lattice is more apparent in the color map shown in the right inset of Fig.~\ref{fig2}(a), which displays $\langle I_{\omega_{\rm L}}(\bold{k}_\parallel)\rangle_{\rm ss}$ across the entire 1BZ. 

Figure~\ref{fig2}(b) displays the total intensity radiated by the QE lattice, $\langle I_\omega \rangle_{\rm ss}=\int_{k_\parallel < k}\langle I_{\omega}(\bold{k}_\parallel) \rangle_{\rm ss}\mathrm{d}\bold{k}_\parallel$, normalized to $I_{\rm L}=P_{\rm L}/l^2$, as a function of $\Omega$ for $\Delta=0$. The green and purple curves correspond to frequency windows centered at  $\omega_{\rm L}$ and $\omega_{\pm}$, respectively, with $\omega_{\pm}$ denoting the Mollow sidebands (see the inset in Fig.~\ref{fig2}(b)).  Again, we distinguish the coherent (orange) and incoherent (blue) contributions to $\langle I_{\omega_{\rm L}}\rangle_{\rm ss}$ while $\langle I_{\omega_{\pm}}\rangle_{\rm ss}$ is entirely incoherent. On the contrary, the corresponding intensity for a CE lattice (black) is completely coherent. The intensity dependence on the driving rate closely follows the trends observed for $\langle N\rangle_{\rm ss}$ in Fig.~\ref{fig1}(b). At strong driving, diffraction effects in the QE lattice are suppressed, and its incoherent emission reduces to $\langle I_{\omega}\rangle_{\rm ss} \approx \int_{\omega}\omega^4 \mu^2/(12\pi\varepsilon_0 c^3 l^2)\mathcal{S}^{\rm I}_{\rm QE, eff}(\omega) \mathrm{d}\omega$, recovering the super-atom radiation limit \cite{Parmee21_PRA, Ruostekoski23_PRA}, but modulated by the incoherent part of the power spectrum. These results are intrinsically linked to the excitation of BSs with parallel wavevectors differing from that of the incident field.

\begin{figure}[t!]
\centering
\includegraphics[width=0.9\columnwidth]{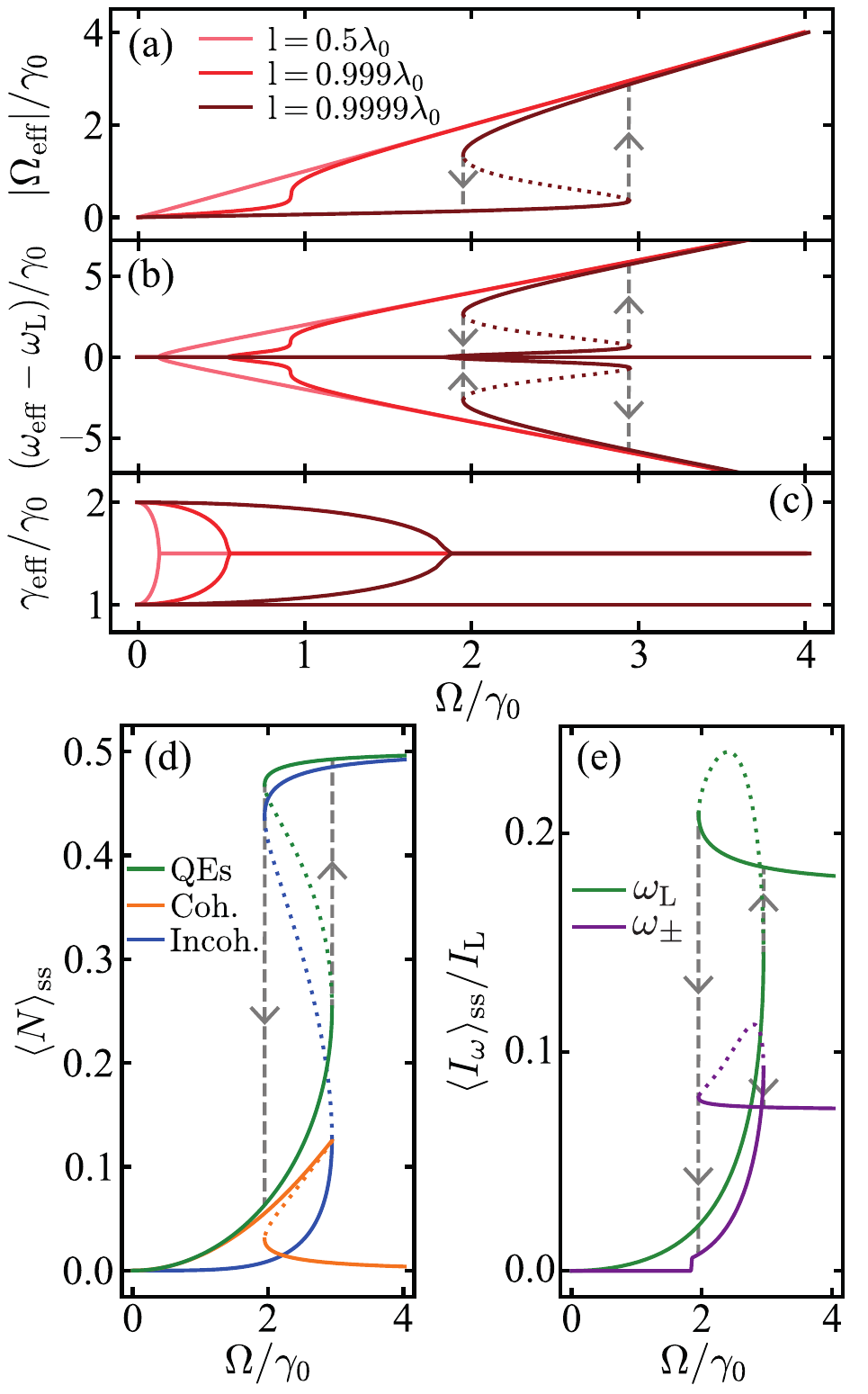}
\caption{Effective driving rates (a), central frequencies (b), and linewidths (c) of the incoherent emission spectrum $\mathcal{S}^{\rm I}_{\rm QE, eff}(\omega)$, as function of $\Omega$ for different values of $l$. (d) Population per emitter, along with its coherent and incoherent contributions, as a function of $\Omega$ for $l=0.9999\lambda_0$. (e) Total intensity emitted by the QE lattice for different frequency windows and $l=0.9999\lambda_0$. Arrows and dotted curves indicate the abrupt transitions and metastable states of the system, respectively ($\Delta =0$).}
\label{fig3}
\end{figure}

To clarify the MFA renormalization of the incoherent emission spectrum from the QE lattice, we examine how the parameters in Eq.~\eqref{Si} depend on $\Omega$ and $l$. As already noted for the configuration with $l=0.5\lambda_0$ considered so far, $\mathcal{S}^{\rm I}_{\rm QE, eff}(\omega)$ resembles that of a single QE. Figure~\ref{fig3}(a) shows the effective driving rate, $\Omega_{\rm eff}$ as a function of $\Omega$ for three different lattice periods, assuming $\Delta = 0$ in all cases. For $l=0.5\lambda_0$, $\Omega_{\rm eff}$ matches $\Omega$, indicating negligible inter-emitter coupling. As the $l$ increases, emitter-emitter interactions become weaker. However, when the period approaches the $\lambda_0$, a bistability in $\Omega_{\rm eff}$ emerges. This effect is particularly evident at $l=0.9999\lambda_0$, where a range of $\Omega$ values yields three possible solutions for the effective driving rate. Two of these are stable, while the third, indicated by a dotted curve, is metastable \cite{Parmee21_PRA}. This hysteresis behavior arises from two factors: the intrinsic anharmonicity of the QEs and the divergence of the lattice sum $\mathcal{G}(\bold{k}_{\rm L, \parallel},\omega_0)$ as $\lambda_0 \rightarrow l$. The latter reflects the emergence of collective, long-range interactions within the QE lattice.

In Figs.~\ref{fig3}(b) and \ref{fig3}(c), we plot the emission frequencies and linewidths, $\omega_{\rm{eff},p}$ and $\gamma_{\rm{eff},p}$ from Eq.~\eqref{Si} ($\Delta = 0$). For $l=0.5\lambda_0$, we observe the characteristic Mollow triplet splitting with increasing driving, and a rapid collapse of all linewidths to $1.5\gamma_0$, matching the single QE behavior. However, when  $l$ approaches $\lambda_0$,  the slower growth of $\Omega_{\rm eff}$ with $\Omega$ delays the onset of resonance fluorescence features, such as sideband emission and linewidth collapse. In this regime, once $\gamma_{\rm{eff},p}=1.5\gamma_0$ for all the Lorentzian peaks, the hysteresis in $\Omega_{\rm eff}$ becomes evident in $\omega_{\rm{eff},p}$, leading to an abrupt transition in the emission spectrum from a singlet to a triplet structure.

To further investigate the impact of the nonlinear behavior of $\Omega_{\rm eff}$ on the optical response of the QE lattice, Figs.~\ref{fig3}(d) and \ref{fig3}(e) show $\langle N \rangle_{\rm ss}$ and $\langle I_\omega\rangle_{\rm ss}$, respectively, as functions of $\Omega$ for $l=0.9999\lambda_0$. The population per emitter reveals that, prior to the onset of bistability, BSs with $\bold{k}_{\parallel}=\bold{k}_{\rm L, \parallel}=0$ are coherently populated (orange), with negligible incoherent contributions (blue) at other wavevectors. As $\Omega$ increases, incoherent population rises sharply only at the upper end of the hysteresis window, whereas for decreasing driving, it dominates $\langle N\rangle_{\rm ss}$ throughout this regime. A similar trend is observed in Fig.~\ref{fig3}(e), which demonstrates that all the emission takes place through $\langle I_{\omega_{\rm L}}\rangle_{\rm ss}$ for low driving. Within the bistability region, $\langle I_{\omega_{\pm}}\rangle_{\rm ss}$ increases smoothly with $\Omega$, reaching a maximum at the upper end of the hysteresis window. For decreasing $\Omega$ it remains nearly constant at this maximum, dropping only at the lower end of the hysteresis.

In conclusion, we have investigated the nonlinear optical response of periodic lattices of quantum emitters under coherent driving. We have demonstrated that in these systems, contrary to their classical counterpart, the external drive does not only populate exciton Bloch states with wavevector given by the incident laser, but a continuum of wavevector states, even beyond the light cone. This phenomenon arises from the inherent quantum anharmonic character of the emitters, which is also responsible for the emergence of nonlinear effects in the resonance fluorescence phenomenology for the lattice. We have shown that the incoherent excitation of detuned Bloch states translates into an incoherent background of photon emission both in frequency and wavevector, which governs the optical response of the system at strong driving. Finally, we have found that collective, long-range emitter interactions give rise to a regime of bistability and hysteresis. This enables abrupt changes in the exciton population and emission properties of the lattice as a function of the driving rate. We believe our findings represent a step toward unveiling the potential of inherent quantum nonlinearities in lattices of emitters for applications in single-photon storage and quantum information technologies.

\section{Aknowledgements}
This work has been generously supported by MCIN/AEI/10.13039/501100011033/FEDER under projects PID2021-126964OB-I00, TED2021-130552B-C21, and PID2022-137569NB-C42. BDA and AIFD acknowledge support from the European Union's Horizon Program through grant 101070700. BDA also thanks the CAM Consejería de Educación, Ciencia y Universidades, Viceconsejería de Universidades, Investigación y Ciencia, Dirección General de Investigación e Innovación Tecnológica (CAM FPI Grant Ref. PIPF-2023/-TEC-29700).

\onecolumngrid
\appendix
\section{Appendix}\label{ap}

\subsection{Lindblad master equation, collective energies and decay rates}
The system under study is a coherently driven periodic lattice of emitters, modeled as point dipoles, which can be quantum or classical.  We represent the quantum (classical) emitters as two-level systems (bosonic resonators) with lower operator $\sigma_i$ ($a_i$). We work within the Markovian approximation in which we restrict ourselves to the weak-coupling regime between the emitters and the photonic free-space bath. The quantum dynamics of the system are given by the following Lindblad master equation
\begin{equation}
    \dot{\rho}={\rm i}\frac{1}{\hbar}[ \rho, H ]+\sum_{i,j}\frac{\gamma_{ij}}{2}L_{\sigma_i,\sigma_j}(\rho),
\label{lindblad first}
\end{equation}
where $H$ is the Hamiltonian of the lattice
\begin{equation}
    H/\hbar=-\Delta\sum_i \sigma_i^\dagger \sigma_i + \sum_i \sum_{j\neq i} g_{ij} \sigma_i^\dagger \sigma_j + \sum_i \left[\Omega e^{{\rm i} \bold{k}_{\rm L, \parallel}\cdot \bold{r}_i} \sigma_i^\dagger+\Omega^* e^{-{\rm i} \bold{k}_{\rm L, \parallel}\cdot \bold{r}_i} \sigma_i \right]
\label{hamiltonian space}
\end{equation}
in the rotating frame of the laser and within the rotating wave approximation. $L_{\sigma_i, \sigma_j}(\rho)=2\sigma_j\rho \sigma^\dagger_i -\{\sigma^\dagger_i \sigma_j, \rho\}$ are Lindblad superoperators, $\Delta=\omega_{\rm L} - \omega_0$ is the detuning between the laser and the natural frequency of the emitter, $\Omega=\bold{E_{\rm L}}\cdot \boldsymbol{\mu} /\hbar$ is the driving amplitude, $\bold{k}_{\rm L, \parallel}$ the parallel component of the wavevector of the laser, 
$\boldsymbol{\mu}$ is the transition dipole of the emitters ($\hat{\boldsymbol{\mu}}=\boldsymbol{\mu}/\mu$), and $\bold{r}_i$  their positions. Moreover, $g_{ij}$ and $\gamma_{ij}$ are the coherent and dissipative interactions, which are given, in terms of the electromagnetic dyadic Green tensor in free space, $\bold{G}(|\bold{r}-\bold{r}_0|,\omega_0),$ by \cite{Dung02_PRA, Novotny12_book}
\begin{align}\nonumber
    &g_{ij}=-\frac{\omega_0^2 \mu^2}{\hbar\varepsilon_0 c^2}\hat{\boldsymbol{\mu}}\cdot \mathrm{Re}\bold{G}(|\bold{r}_i-\bold{r}_j|,\omega_0)\cdot\hat{\boldsymbol{\mu}},\\
    &\gamma_{ij}=\frac{2\omega_0^2 \mu^2}{\hbar\varepsilon_0 c^2}\hat{\boldsymbol{\mu}}\cdot \mathrm{Im}\bold{G}(|\bold{r}_i-\bold{r}_j|,\omega_0)\cdot\hat{\boldsymbol{\mu}},
    \label{y_ij}
\end{align}
with $\gamma_0=\omega_0^3 \mu^2/(3 \varepsilon_0 \hbar \pi c^3)$ being the emitter decay rate in free space, which can be calculated as $\gamma_{ii}$ in Equation~\eqref{y_ij}. The undriven Hamiltonian can be diagonalized in BSs with a well-defined parallel wavevector that are generated by the creation operator $\sigma^\dagger_{\bold{k}_\parallel}=l/(2\pi) \sum_i e^{{\rm i} \bold{k}_\parallel \cdot \bold{r}_i}\sigma^\dagger_i$, where $l$ is the period of the lattice and $k=\omega/c$. Equation~\eqref{lindblad first}, rewritten in terms of $\sigma_{\bold{k}_\parallel}$, reads
\begin{equation*}
    H/\hbar=\int_{\rm 1BZ}\mathrm{d}\bold{k}_\parallel \left(\omega_0 - \omega_{\rm L} - \frac{\omega_0^2 \mu^2}{\hbar\varepsilon_0c^2}\mathrm{Re}\mathcal{G}(\bold{k}_\parallel,\omega_0)\right)\sigma^\dagger_{\bold{k}_\parallel}\sigma_{\bold{k}_\parallel} + \frac{2\pi}{l}\left( \Omega \sigma^\dagger_{\bold{k}_{\rm L, \parallel}} + \Omega^* \sigma_{\bold{k}_{\rm L, \parallel}} \right).
\end{equation*}
In the same way, the Lindblad superoperators can be rewritten as
\begin{equation*}
   \sum_{i,j}\frac{\gamma_{ij}}{2}L_{\sigma_i,\sigma_j}(\rho)=\int_{\rm 1BZ}\mathrm{d}\bold{k}_\parallel \left(\frac{\gamma_0}{2}+\frac{\omega_0^2 \mu^2}{\hbar\varepsilon_0c^2}\mathrm{Im}\mathcal{G}(\bold{k}_\parallel,\omega_0)\right)L_{\sigma_{\bold{k}_\parallel}}(\rho),
\end{equation*}
where $L_{\sigma_{\bold{k}_\parallel}}(\rho)=2\sigma_{\bold{k}_\parallel}\rho \sigma^\dagger_{\bold{k}_\parallel} - \{\sigma^\dagger_{\bold{k}_\parallel}\sigma_{\bold{k}_\parallel},\rho\}$. Here, $\mathcal{G}(\bold{k}_\parallel,\omega_0)=\sum_{j\neq 0}\hat{\boldsymbol{\mu}}\cdot\bold{G}(|\bold{r}_j|,\omega_0)\cdot\hat{\boldsymbol{\mu}}e^{-i\bold{k}_\parallel\cdot \bold{r}_j}$ is commonly referred as the lattice sum and can be efficiently computed using Ewald's method \cite{Kambe68_ZA}. The eigenenergies and decay rates of the BSs are $\hbar\Delta_{\bold{k}_\parallel}=\hbar\Delta + \tfrac{\omega_0^2 \mu^2}{\varepsilon_0 c^2}\mathrm{Re} \mathcal{G}(\bold{k}_\parallel, \omega_0)$ and $\gamma_{\bold{k}_{\parallel}}=\gamma_0 + \tfrac{2\omega_0^2 \mu^2}{\hbar \varepsilon_0 c^2}\mathrm{Im} \mathcal{G}(\bold{k}_{\parallel},\omega_0)$, while the laser driving only acts, directly, on the state with $\bold{k}_\parallel = \bold{k}_{\rm L,\parallel}$. An example of the behavior of these parameters in 1D, 2D, and 3D can be seen in  Ref.~\cite{Asenjo17_PRX}. These results are equivalent for QEs and CEs since, in the single-excitation subspace, the nonlinear character of the emitters does not play any role. However, we demonstrate in the next sections that, beyond the single excitation limit, the BS populations and the radiated intensity are different depending on whether the lattice is composed by classical or quantum emitters. All of the calculations that we present in what follows are performed in the steady-state regime.

\subsection{Results For a Lattice of Classical Emitters}
In this Section, we compute the steady-state correlator $\langle a^\dagger_{\bold{k}_\parallel}a_{\bold{k}_\parallel^\prime} \rangle_{\rm ss}$ as well as the steady-state power spectrum defined as $\mathcal{S}_{\bold{k}_\parallel,\bold{k}^\prime_\parallel}(\omega)=\lim_{t\rightarrow \infty}\tfrac{1}{\pi}\mathrm{Re}\int_{0}^{\infty}\langle a^\dagger_{\bold{k}_\parallel}
(t)a_{\bold{k}^\prime_\parallel}(t+\tau)\rangle e^{i\omega \tau}\mathrm{d}\tau=\tfrac{1}{\pi}\mathrm{Re}\int_{0}^{\infty}\langle a^\dagger_{\bold{k}_\parallel}
(0)a_{\bold{k}^\prime_\parallel}(\tau)\rangle_{\rm ss}{} e^{i\omega \tau}\mathrm{d}\tau$. We start with the simpler case of a lattice of CEs. The emitters are described by bosonic operators, $a_i$, which follow the canonical commutation relations $[a_i,a^\dagger_j]=\delta_{ij}$ and $[a_i,a_j]=0$. Therefore, the commutation relations of the Bloch operators $a_{\bold{k}_\parallel}$ are $[a_{\bold{k}_\parallel},a^\dagger_{\bold{k}^\prime_\parallel}]=\delta(\bold{k}_\parallel - \bold{k}^\prime_\parallel)$ and $[a_{\bold{k}_\parallel},a_{\bold{k}^\prime_\parallel}]=0$. The dynamics of $\langle a_{\bold{k}_\parallel} \rangle$ can be computed as
\begin{align}\nonumber
    \frac{\rm d}{{\rm d}t}\langle a_{\bold{k}_\parallel} \rangle&={\rm Tr}\{a_{\bold{k}_\parallel} \dot{\rho}\}=\\ \nonumber
    &=\int_{\rm 1BZ} {\rm Tr}\left\{a_{\bold{k}_\parallel}\left({\rm i}[\rho, -\Delta_{\bold{k}^\prime_\parallel}a^\dagger_{\bold{k}^\prime_\parallel}a_{\bold{k}^\prime_\parallel}] + \frac{\gamma_{\bold{k}^\prime_\parallel}}{2}L_{a_{\bold{k}^\prime_\parallel}}(\rho)\right)\right\}\mathrm{d}\bold{k}_\parallel^\prime + {\rm i}\frac{2\pi}{l}{\rm Tr}\left\{a_{\bold{k}_\parallel}[\rho,\Omega a^\dagger_{\bold{k}_{\rm L, \parallel}} + \Omega^* a_{\bold{k}_{\rm L, \parallel}}]\right\}=\\
    &=\left({\rm i}\Delta_{\bold{k}_\parallel} -\frac{\gamma_{\bold{k}_\parallel}}{2}\right)\langle a_{\bold{k}_\parallel} \rangle -{\rm i}\Omega\frac{2\pi}{l}\delta(\bold{k}_\parallel - \bold{k}_{\rm L, \parallel}).
\label{dynamics of ak}
\end{align}
In the steady-state ($\tfrac{\rm d}{{\rm d}t}\langle a_{\bold{k}_\parallel} \rangle_{\rm ss}=0)$, Equation~\eqref{dynamics of ak} reduces to
\begin{equation}
    \langle a_{\bold{k}_\parallel} \rangle_{\rm ss}=\frac{\Omega}{\Delta_{\bold{k}_\parallel} + {\rm i}\gamma_{\bold{k}_\parallel}/2}\frac{2\pi}{l}\delta(\bold{k}_\parallel - \bold{k}_{\rm L, \parallel}).
    \label{a k exact}
\end{equation}
Now, in order to calculate the power spectrum, we need to compute the two-times correlator $\langle a^\dagger_{\bold{k}_\parallel}
(t)a_{\bold{k}^\prime_\parallel}(t+\tau)\rangle$. To that end, we use the Quantum Regression Theorem (QRT) \cite{Meystre07_book, Blocher19_PRA}. Mathematically, the QRT states that once we find the system of differential equations for $\bold{c}_{\bold{k}_\parallel}(t)$, where $\bold{c}_{\bold{k}_\parallel}(t)$ is a set of coupled operators, such that ${\rm d}\langle \bold{c}_{\bold{k}_\parallel}(t)\rangle/{\rm d}t=M\langle\bold{c}_{\bold{k}_\parallel}(t)\rangle + \bold{b}$, then
\begin{equation}
    \frac{\rm d}{{\rm d}\tau}\langle O_1(t)\bold{c}_{\bold{k}_\parallel}(t+\tau)O_2(t)\rangle=M\langle O_1(t)\bold{c}_{\bold{k}_\parallel}(t+\tau)O_2(t)\rangle + \langle O_1(t)O_2(t)\rangle \bold{b},
\label{qrt}
\end{equation}
where $O_1$ and $O_2$ are any two operators of the system. Here, we are interested in the two-times correlator $\langle a_{\bold{k}_\parallel}^\dagger(t)a_{\bold{k}_\parallel^\prime}(t+\tau) \rangle$ so, substituting $O_1(t)=a_{\bold{k}_\parallel}^\dagger(t)$ and $O_2(t)=1$ into Equation~\eqref{qrt}, we get
\begin{equation*}
    \frac{\rm d}{{\rm d}\tau}\langle a_{\bold{k}_\parallel}^\dagger(t)a_{\bold{k}_\parallel^\prime}(t+\tau)\rangle=\left( {\rm i}\Delta_{\bold{k}^\prime_\parallel} - \frac{\gamma_{\bold{k}^\prime_\parallel}}{2}\right)\langle a_{\bold{k}_\parallel}^\dagger(t)a_{\bold{k}_\parallel^\prime}(t+\tau)\rangle - {\rm i}\Omega\langle a_{\bold{k}_\parallel}^\dagger(t)\rangle\frac{2\pi}{l}\delta(\bold{k}^\prime_\parallel - \bold{k}_{\rm L, \parallel}).
\end{equation*}
The solution of this differential equation for $t \rightarrow \infty$ is
\begin{align*}\nonumber
     \langle a_{\bold{k}_\parallel}^\dagger(0)a_{\bold{k}_\parallel^\prime}(\tau)\rangle_{\rm ss} = {}& \left( \langle a^\dagger_{\bold{k}_\parallel}a_{\bold{k}^\prime_\parallel}\rangle_{\rm ss} - \frac{|\Omega|^2}{\Delta_{\bold{k}_{\rm L, \parallel}}^2 + \gamma_{\bold{k}_{\rm L, \parallel}}^2/4}\frac{4\pi^2}{l^2}\delta(\bold{k}_\parallel - \bold{k}_{\rm L, \parallel})\delta(\bold{k}^\prime_\parallel - \bold{k}_{\rm L, \parallel})  \right)e^{({\rm i}\Delta_{\bold{k}^\prime_\parallel}-\gamma_{\bold{k}^\prime_\parallel}/2)\tau} + \\ 
    &+ \frac{|\Omega|^2}{\Delta_{\bold{k}_{\rm L, \parallel}}^2 + \gamma_{\bold{k}_{\rm L, \parallel}}^2/4}\frac{4\pi^2}{l^2}\delta(\bold{k}_\parallel - \bold{k}_{\rm L, \parallel})\delta(\bold{k}^\prime_\parallel - \bold{k}_{\rm L, \parallel}),
\end{align*}
and, therefore, the power spectrum is 
\begin{align*}\nonumber
    \mathcal{S}_{\bold{k}_\parallel,\bold{k}^\prime_\parallel}(\omega)={}&\frac{1}{\pi}\left( \langle a^\dagger_{\bold{k}_\parallel}a_{\bold{k}^\prime_\parallel}\rangle_{\rm ss} - \frac{|\Omega|^2}{\Delta_{\bold{k}_{\rm L, \parallel}}^2 + \gamma_{\bold{k}_{\rm L, \parallel}}^2/4}\frac{4\pi^2}{l^2}\delta(\bold{k}_\parallel - \bold{k}_{\rm L, \parallel})\delta(\bold{k}^\prime_\parallel - \bold{k}_{\rm L, \parallel})  \right)\frac{\gamma_{\bold{k}^\prime_\parallel}/2}{(\omega + \Delta_{\bold{k}^\prime_\parallel})^2 + \gamma_{\bold{k}^\prime_\parallel}^2/4}+\\
    &+ \frac{|\Omega|^2}{\Delta_{\bold{k}_{\rm L, \parallel}}^2 + \gamma_{\bold{k}_{\rm L, \parallel}}^2/4}\frac{4\pi^2}{l^2}\delta(\bold{k}_\parallel - \bold{k}_{\rm L, \parallel})\delta(\bold{k}^\prime_\parallel - \bold{k}_{\rm L, \parallel})\delta(\omega).
    \label{spectrum bosons casi}
\end{align*}
To complete the calculation of $ \mathcal{S}_{\bold{k}_\parallel,\bold{k}^\prime_\parallel}(\omega)$, we also need to compute the correlator $\langle a^\dagger_{\bold{k}_\parallel}a_{\bold{k}^\prime_\parallel}\rangle_{\rm ss}$. In a similar way to Equation~\eqref{dynamics of ak} we get
\begin{equation}
    \langle a^\dagger_{\bold{k}_\parallel}a_{\bold{k}^\prime_\parallel}\rangle_{\rm ss}=\frac{4\pi^2}{l^2}\frac{|\Omega|^2}{\Delta^2_{\bold{k}_{\rm L, \parallel}}+\gamma^2_{\bold{k}_{\rm L, \parallel}}/4}\delta(\bold{k}_\parallel - \bold{k}_{\rm L, \parallel})\delta(\bold{k}^\prime_\parallel - \bold{k}_{\rm L, \parallel}).
\label{Pop as}
\end{equation}
where, in the last identity, we have used the previous results of $\langle a_{\bold{k}_\parallel} \rangle_{\rm ss}$ and $\langle a^\dagger_{\bold{k}_\parallel} \rangle_{\rm ss}$. Equation~\eqref{Pop as} implies both that $\langle a^\dagger_{\bold{k}_\parallel}a_{\bold{k}^\prime_\parallel}\rangle_{\rm ss}$ is $0$ if $\bold{k}_\parallel\neq \bold{k}^\prime_\parallel$ and that the only populated state is the one with $\bold{k}_\parallel=\bold{k}_{\rm L, \parallel}$. This can easily be seen in the resulting population distribution inside the First Brillouin Zone ($\rm 1BZ$)
\begin{equation}
    \langle n_{\bold{k}_\parallel} \rangle_{\rm ss}=\frac{l}{2\pi}\int_{\rm 1BZ}\langle a^\dagger_{\bold{k}_\parallel}a_{\bold{k}_\parallel^\prime} \rangle_{\rm ss} \mathrm{d}\bold{k}_\parallel^\prime = \frac{|\Omega|^2}{\Delta^2_{\bold{k}_{\rm L,\parallel}} + \gamma^2_{\bold{k}_{\rm L,\parallel}}/4} \frac{2\pi}{l}\delta(\bold{k}_\parallel - \bold{k}_{\rm L,\parallel}),
\label{n a}
\end{equation}
which indicates that the distribution is just given by a Dirac delta centered at $\bold{k}_{\rm L,\parallel}$. In addition, $\langle a^\dagger_{\bold{k}_\parallel}a_{\bold{k}^\prime_\parallel}\rangle_{\rm ss}=\langle a^\dagger_{\bold{k}_\parallel}\rangle_{\rm ss} \langle a_{\bold{k}^\prime_\parallel}\rangle_{\rm ss}$, implying that the operators $a_{\bold{k}_\parallel}$ are not quantum correlated. Finally, after returning the system to the atomic frame, the spectrum is
\begin{align}
  \mathcal{S}_{\bold{k}_\parallel,\bold{k}^\prime_\parallel}(\omega)= \frac{|\Omega|^2}{\Delta_{\bold{k}_{\rm L,\parallel}}^2 + \gamma_{\bold{k}_{\rm L, \parallel}}^2/4}\frac{4\pi^2}{l^2}\delta(\bold{k}_\parallel - \bold{k}_{\rm L, \parallel})\delta(\bold{k}_\parallel^\prime - \bold{k}_{\rm L, \parallel})\delta(\omega-\omega_{\rm L}).
\label{spectrum bosons}
\end{align}
Equations~\eqref{Pop as},~\eqref{n a} and~\eqref{spectrum bosons} show that a lattice of CEs is a linear system for which the emitted photons have the same wavevector and frequency than the incident ones and therefore momentum and energy are trivially conserved.

\subsection{Results for a Lattice of Quantum Emitters}
Next, we calculate the same quantities for a lattice of QEs, revealing a more intricate problem structure. First of all, taking into account the commutation relations of the QEs, $\{\sigma_i,\sigma^\dagger_i\}=1$ and $[\sigma_i,\sigma_j]=[\sigma_i,\sigma^\dagger_j]=0$ for $i\neq j$ or, in a more compacted form, $[\sigma_i,\sigma^\dagger_j]=\delta_{ij}(1-2\sigma^\dagger_i\sigma_i)$ and $[\sigma_i,\sigma_j]=0$, we readily find that the corresponding $\sigma_{\bold{k}_\parallel}$ operators commute as $[\sigma_{\bold{k}_\parallel},\sigma_{\bold{k}^\prime_\parallel}]=0$ and
\begin{equation*}
    [\sigma_{\bold{k}_\parallel},\sigma^\dagger_{\bold{k}^\prime_\parallel}]=\delta(\bold{k}_\parallel - \bold{k}^\prime_\parallel) - 2\mathcal{O}_{\bold{k}_\parallel, \bold{k}^\prime_\parallel},
\end{equation*}
where we define
\begin{equation*}
    \mathcal{O}_{\bold{k}_\parallel, \bold{k}^\prime_\parallel}=\frac{l^2}{4\pi^2}\sum_{i}e^{-{\rm i}\bold{k}_\parallel \cdot \bold{r}_i}e^{i\bold{k}^\prime_\parallel \cdot \bold{r}_i}\sigma_i^\dagger\sigma_i.
\end{equation*}
The operators $\sigma_{\bold{k}_\parallel}$ do not have a simple commutation relation (bosonic or fermionic) but follow more complex rules due to the spin-algebra of $\sigma_i$. However, although these operators do not commute, the states $\ket{\bold{k}_\parallel}=\sigma^\dagger_{\bold{k}_\parallel}\ket{0}$ are still orthogonal. Importantly, among other properties, the operator $\mathcal{O}_{\bold{k}_\parallel, \bold{k}_\parallel^\prime}$ satisfies that when applied to a certain state $\ket{\bold{k}^{\prime \prime}_\parallel}$ it returns $\mathcal{O}_{\bold{k}_\parallel, \bold{k}^\prime_\parallel}\ket{\bold{k}^{\prime \prime}_\parallel}=l^2/(4\pi^2)\ket{\bold{k}^{\prime \prime}_\parallel + \bold{k}^{\prime}_\parallel - \bold{k}_\parallel}$, inducing correlations between different BSs. Note that these correlations are not due to the interactions between the emitters but just by their algebra, which differs from the classical case due to the quantum nonlinearity 
inherent to the QEs.

As before, we need to compute $\langle \sigma_{\bold{k}_\parallel} \rangle_{\rm ss}$. The corresponding differential equation can be written as
\begin{align}\nonumber
    \frac{\rm d}{{\rm d}t}\langle \sigma^\dagger_{\bold{k}_\parallel}\rangle={}&
    \left({\rm i}\Delta_{\bold{k}_\parallel}-\frac{\gamma_{\bold{k}_\parallel}}{2}\right)\langle \sigma_{\bold{k}_\parallel}\rangle -2\int_{\rm 1BZ}\mathrm{d}\bold{k}_\parallel^\prime \left({\rm i}\Delta_{\bold{k}^\prime_\parallel}-\frac{\gamma_{\bold{k}^\prime_\parallel}}{2}\right)\langle \mathcal{O}_{\bold{k}_\parallel,\bold{k}^\prime_\parallel}\sigma_{\bold{k}^\prime_\parallel}\rangle\\
    &-{\rm i}\Omega\frac{2\pi}{l}\delta(\bold{k}_\parallel - \bold{k}_{\rm L, \parallel}) + 2{\rm i}\Omega\frac{2\pi}{l}\langle \mathcal{O}_{\bold{k}_\parallel,\bold{k}_{\rm L, \parallel}}\rangle. \label{sigmadagger}
\end{align}
The equation above includes the expected values of $\mathcal{O}_{\bold{k}_\parallel,\bold{k}^\prime_\parallel}$ and $ \mathcal{O}_{\bold{k}_\parallel,\bold{k}^\prime_\parallel}\sigma_{\bold{k}^\prime_\parallel}$, which take into account the quantum features of the QEs. Without the contributions of these correlators the system would behave linearly as the CE lattice. When we compute the equations for the evolution of these expected values, we find that they depend on the expected values of more complicated correlators involving $\mathcal{O}_{\bold{k}_\parallel,\bold{k}^\prime_\parallel}$ and $\sigma_{\bold{k}_\parallel}$. This makes impossible to obtain a closed expression for the BS population and power spectrum. To circumvent this problem, we neglect the interactions between the emitters, setting $\Delta_{\bold{k}_\parallel}\approx\Delta$ and $\gamma_{\bold{k}_\parallel}\approx\gamma_0$ (although we later include these interactions in Section IV using a MFA). With this approximation,
\begin{align*}
    \int_{\rm 1BZ}\mathrm{d}\bold{k}^\prime_\parallel \langle \mathcal{O}_{\bold{k}_\parallel,\bold{k}^\prime_\parallel}\sigma_{\bold{k}^\prime_\parallel}\rangle = \frac{l^2}{4\pi^2}\frac{l}{2\pi}\sum_{i,j}e^{-{\rm i}\bold{k}_\parallel\cdot \bold{r}_i}\int_{\rm 1BZ}\mathrm{d}\bold{k}_\parallel^\prime e^{{\rm i}\bold{k}^\prime_\parallel \cdot(\bold{r}_i - \bold{r}_j)}\langle \sigma^\dagger_i\sigma_i\sigma_j\rangle=\frac{l}{2\pi}\sum_{i,j}e^{-{\rm i}\bold{k}_\parallel\cdot \bold{r}_i} \delta_{ij}\langle \sigma^\dagger_i\sigma_i\sigma_j\rangle=0.
\end{align*}
Furthermore, if we calculate ${\rm d}\langle \mathcal{O}_{\bold{k}_\parallel,\bold{k}_{\rm L, \parallel}} \rangle/{\rm d}t$, we get 
\begin{align*}
    &\frac{\rm d}{{\rm d}t}\langle \mathcal{O}_{\bold{k}_\parallel,\bold{k}_{\rm L, \parallel}} \rangle=-\gamma_0 \langle \mathcal{O}_{\bold{k}_\parallel,\bold{k}_{\rm L, \parallel}} \rangle -{\rm i}\frac{l}{2\pi}\Omega \langle \sigma^\dagger_{2\bold{k}_{\rm L, \parallel} - \bold{k}_\parallel}\rangle + {\rm i}\frac{l}{2\pi}\Omega^* \langle \sigma_{\bold{k}_\parallel}\rangle,
\end{align*}
so the evolution of $\langle \mathcal{O}_{\bold{k}_\parallel,\bold{k}_{\rm L, \parallel}} \rangle$ depends on $\langle \sigma^\dagger_{2\bold{k}_{\rm L, \parallel} - \bold{k}_\parallel}\rangle$ and on $\langle \sigma_{\bold{k}_\parallel}\rangle$. We can obtain ${\rm d}\langle \sigma^\dagger_{2\bold{k}_{\rm L, \parallel} - \bold{k}_\parallel}\rangle/{\rm d}t$ from Equation~\eqref{sigmadagger} by substituting $\bold{k}_\parallel$ with $2\bold{k}_{\rm L, \parallel} - \bold{k}_\parallel$
\begin{align*}
    \frac{\rm d}{{\rm d}t}\langle \sigma^\dagger_{2\bold{k}_{\rm L, \parallel} - \bold{k}_\parallel}\rangle=\left(-{\rm i}\Delta - \frac{\gamma_0}{2}\right)\langle \sigma^\dagger_{2\bold{k}_{\rm L, \parallel} - \bold{k}_\parallel}\rangle + {\rm i}\Omega^*\frac{2\pi}{l}\delta(\bold{k}_\parallel - \bold{k}_{\rm L, \parallel}) - {\rm i}2\Omega^*\frac{2\pi}{l}\langle \mathcal{O}_{\bold{k}_\parallel,\bold{k}_{\rm L, \parallel}} \rangle,
\end{align*}
where we have used $\mathcal{O}^\dagger_{2\bold{k}_{\rm L, \parallel}-\bold{k}_\parallel, \bold{k}_{\rm L, \parallel}}=\mathcal{O}_{\bold{k}_\parallel,\bold{k}_{\rm L, \parallel}}$. Note that the chain of coupled correlators ends here, showing that $\langle \sigma_{\bold{k}_\parallel}\rangle$, $\langle \sigma^\dagger_{2\bold{k}_{\rm L, \parallel} - \bold{k}_\parallel}\rangle$ and $\langle \mathcal{O}_{\bold{k}_\parallel,\bold{k}_{\rm L, \parallel}} \rangle$ form a closed linear system of differential equations. In a matrix form, the system of differential equations can be written as
\begin{align*}\nonumber
    \frac{\rm d}{{\rm d}t}\begin{pmatrix}
                \langle \sigma_{\bold{k}_{\parallel}} \rangle\\
                \langle \sigma^\dagger_{2\bold{k}_{\rm L, \parallel}-\bold{k}_\parallel} \rangle\\
                \langle \frac{2\pi}{l}\mathcal{O}_{\bold{k}_\parallel,\bold{k}_{\rm L, \parallel}} \rangle
                \end{pmatrix}
=&\begin{pmatrix}
    {\rm i}\Delta-\frac{\gamma_0}{2} & 0 & {\rm i}2\Omega\\
    0 & -{\rm i}\Delta-\frac{\gamma_0}{2} & -{\rm i}2\Omega^*\\
    {\rm i}\Omega^* & -{\rm i}\Omega & -\gamma_0
\end{pmatrix}
\begin{pmatrix}
               \langle \sigma_{\bold{k}_{\parallel}} \rangle\\
                \langle \sigma^\dagger_{2\bold{k}_{\rm L, \parallel}-\bold{k}_\parallel} \rangle\\
                \langle \frac{2\pi}{l}\mathcal{O}_{\bold{k}_\parallel,\bold{k}_{\rm L, \parallel}} \rangle
                \end{pmatrix}+\\
&+ \begin{pmatrix}
    -\Omega\\
    \Omega^*\\
    0
\end{pmatrix}
{\rm i}\frac{2\pi}{l}\delta(\bold{k}_\parallel-\bold{k}_{\rm L, \parallel})=M\langle \bold{c}_{\bold{k}_\parallel} \rangle + \bold{b},
\label{evol. Equation}
\end{align*}
where $\langle \bold{c}_{\bold{k}_\parallel} \rangle=(\langle \sigma_{\bold{k}_{\parallel}} \rangle, \langle \sigma^\dagger_{2\bold{k}_{\rm L, \parallel}-\bold{k}_\parallel} \rangle, \langle \frac{2\pi}{l}\mathcal{O}_{\bold{k}_\parallel,\bold{k}_{\rm L, \parallel}} \rangle)$, ${\bold{b}=(-{\rm i}\Omega,{\rm i}\Omega^*,0)\frac{2\pi}{l}\delta(\bold{k}_\parallel-\bold{k}_{\rm L, \parallel})}$, and
\begin{equation*}
    M=\begin{pmatrix}
    {\rm i}\Delta-\frac{\gamma_0}{2} & 0 & {\rm i}2\Omega\\
    0 & -{\rm i}\Delta-\frac{\gamma_0}{2} & -{\rm i}2\Omega^*\\
    {\rm i}\Omega^* & -{\rm i}\Omega & -\gamma_0
\end{pmatrix} 
\end{equation*}
is the regression matrix. The steady-state solution of $\langle \bold{c}_{\bold{k}_\parallel} \rangle$ is

\begin{equation*}
    \langle \sigma_{\bold{k}_\parallel} \rangle_{\rm ss}=-\frac{{\rm i}2\Omega(\gamma_0 + {\rm i}2\Delta)}{\gamma_0^2 + 4\Delta^2 + 8|\Omega|^2}\frac{2\pi}{l}\delta(\bold{k}_\parallel - \bold{k}_{\rm L, \parallel}),
\end{equation*}
\begin{equation*}
    \langle \sigma^\dagger_{2\bold{k}_{\rm L, \parallel}-\bold{k}_\parallel} \rangle_{\rm ss}=\frac{{\rm i}2\Omega^*(\gamma_0 - {\rm i}2\Delta)}{\gamma_0^2 + 4\Delta^2 + 8|\Omega|^2}\frac{2\pi}{l}\delta(\bold{k}_\parallel - \bold{k}_{\rm L, \parallel}),
\end{equation*}
\begin{equation*}
    \frac{2\pi}{l} \langle\mathcal{O}_{\bold{k}_\parallel,\bold{k}_{\rm L, \parallel}} \rangle_{\rm ss}=\frac{4|\Omega|^2}{\gamma_0^2+4\Delta^2+8|\Omega|^2}\delta(\bold{k}_\parallel - \bold{k}_{\rm L, \parallel}).
\end{equation*}

Applying the QRT, we obtain the following steady-state solution
\begin{equation}
    \langle \sigma_{\bold{k}_\parallel}^\dagger(0)\sigma_{\bold{k}_\parallel^\prime}(\tau)\rangle_{\rm ss}=\sum_{p=1}^{3}l_pe^{\lambda_p\tau}+\langle\sigma_{\bold{k}_\parallel}^\dagger\rangle_{\rm ss}\langle\sigma_{\bold{k}_\parallel^\prime}\rangle_{\rm ss},
\label{eq con lp}
\end{equation}
where $l_p=\sum_{i=1}^{3}E_{1p}E^{-1}_{pi}(\bold{v}_{\rm ss})_i$, with $E_{ij}$ the elements of the matrix of column eigenvectors of $M$, and $\lambda_p$ its p-th eigenvalue. Also, we define $\bold{v}_{\rm ss}=\langle \sigma_{\bold{k}_\parallel}^\dagger\bold{c}_{\bold{k}_\parallel^\prime}\rangle_{\rm ss}-\langle \sigma_{\bold{k}_\parallel}^\dagger\rangle_{\rm ss}\langle\bold{c}_{\bold{k}_\parallel^\prime}\rangle_{\rm ss}$. Next, in order to compute the power spectrum, we Fourier-transform Equation~\eqref{eq con lp} and get
\begin{align}\nonumber
    \mathcal{S}_{\bold{k}_\parallel,\bold{k}_\parallel^\prime}(\omega)={}&\frac{4|\Omega|^2(\gamma_0^2+4\Delta^2)}{(\gamma_0^2+4\Delta^2+8|\Omega|^2)^2}\frac{4\pi^2}{l^2}\delta(\bold{k}_\parallel-\bold{k}_{\rm L, \parallel})\delta(\bold{k}_\parallel^\prime-\bold{k}_{\rm L,\parallel})\delta(\omega-\omega_{\rm L})+\\
    &+\frac{1}{\pi}\sum_{p=1}^{3}\left(L_p \frac{\gamma_p/2}{(\omega-\omega_{\rm L}-\omega_p)^2+(\gamma_p/2)^2} - K_p\frac{\omega-\omega_{\rm L}-\omega_p}{(\omega-\omega_{\rm L}-\omega_p)^2+(\gamma_p/2)^2}\right),
    \label{S casi}
\end{align}
where $\gamma_p=-2\mathrm{Re}(\lambda_p)$, $\omega_p=-\mathrm{Im}(\lambda_p)$, $L_p=\mathrm{Re}(l_p)$ and $K_p=\mathrm{Im}(l_p)$. Similar to the classical case, we need to compute $\bold{v}_{\rm ss}$ to obtain the power spectrum. This calculation involves the correlators $\langle \sigma^\dagger_{\bold{k}_\parallel} \sigma_{\bold{k}^\prime_\parallel}\rangle_{\rm ss}$, $\langle \sigma^\dagger_{\bold{k}_\parallel} \sigma^\dagger_{2\bold{k}_{\rm L, \parallel} - \bold{k}^\prime_\parallel}\rangle_{\rm ss}$, and $\langle \sigma^\dagger_{\bold{k}_\parallel} \mathcal{O}_{\bold{k}^\prime_\parallel,\bold{k}_{\rm L, \parallel}}\rangle_{\rm ss}$. In this case, although they can be calculated as before in the reciprocal space, it is easier to proceed in the position space. The populations and coherences of a lattice of noninteracting QEs in position space are
\begin{equation*}
    \langle \sigma_i^\dagger \sigma_i \rangle_{\rm ss}=\frac{4|\Omega|^2}{\gamma_0^2+4\Delta^2+8|\Omega|^2},
\label{population}
\end{equation*}
\begin{equation}
    \langle \sigma_i^\dagger \sigma_j \rangle_{\rm ss}=\langle \sigma_i^\dagger \rangle_{\rm ss} \langle \sigma_j \rangle_{\rm ss}=\frac{4|\Omega|^2 (\gamma_0^2+ 4\Delta^2)}{(\gamma_0^2+4\Delta^2+8|\Omega|^2)^2}e^{-i\bold{k}_{\rm L, \parallel}\cdot (\bold{r_i}-\bold{r_j})}.
\label{i neq j}
\end{equation}
Importantly, Equation~\eqref{i neq j} only holds for $i\neq j$, while $ \langle \sigma_i^\dagger \sigma_i \rangle_{\rm ss} \neq \langle \sigma_i^\dagger \rangle_{\rm ss} \langle \sigma_i \rangle_{\rm ss}$. The correlator $\langle \sigma^\dagger_{\bold{k}_\parallel} \sigma_{\bold{k}_\parallel^\prime}\rangle_{\rm ss}$ is then
 \begin{equation}
    \langle \sigma_{\bold{k}_\parallel}^\dagger \sigma_{\bold{k}_\parallel^\prime}\rangle_{\rm ss}=\frac{4\pi^2}{l^2}\left[\frac{4|\Omega|^2 (\gamma_0^2+ 4\Delta^2)}{(\gamma_0^2+4\Delta^2+8|\Omega|^2)^2} \right]\delta(\bold{k}_\parallel-\bold{k}_{\rm L, \parallel})\delta(\bold{k}_\parallel^\prime-\bold{k}_{\rm L, \parallel})+ \left[\frac{32|\Omega|^4}{(\gamma_0^2+4\Delta^2+8|\Omega|^2)^2}\right]\delta(\bold{k}_\parallel-\bold{k}_\parallel^\prime).
\label{pops 2LS}
\end{equation}
Equation~\eqref{pops 2LS} implies a nonzero population of BSs with $\bold{k}_\parallel\neq\bold{k}_{\rm L, \parallel}$ and that different BSs are correlated since $\langle \sigma^\dagger_{\bold{k}_\parallel} \sigma_{\bold{k}_\parallel^\prime} \rangle_{\rm ss} \neq \langle \sigma^\dagger_{\bold{k}_\parallel}\rangle_{\rm ss} \langle \sigma_{\bold{k}_\parallel^\prime} \rangle_{\rm ss}$, even though the QEs are not interacting.

The other correlators can be calculated in a similar way
\begin{align}
    &\langle \sigma^\dagger_{\bold{k}_\parallel}\sigma^\dagger_{2\bold{k}_{\rm L, \parallel} - \bold{k}^\prime_\parallel}\rangle_{\rm ss} =\frac{-4(\gamma_0-{\rm i}2\Delta)^2\left(\Omega^*\right)^2}{(\gamma_0^2+4\Delta^2+8|\Omega|^2)^2}\left[\frac{4\pi^2}{l^2}\delta(\bold{k}_\parallel-\bold{k}_{\rm L, \parallel})\delta(\bold{k}_\parallel^\prime-\bold{k}_{\rm L, \parallel}) - \delta(\bold{k}_\parallel - \bold{k}^\prime_\parallel)\right],
\label{sigma dag sigma dag_1}
\end{align}
\begin{align}
    &\frac{2\pi}{l}\langle \sigma^\dagger_{\bold{k}_\parallel}\mathcal{O}_{\bold{k}^\prime_\parallel, \bold{k}_{\rm L, \parallel}}\rangle_{\rm ss}=\frac{{\rm i}8(\gamma_0-{\rm i}2\Delta)\Omega^*|\Omega|^2}{(\gamma_0^2+4\Delta^2+8|\Omega|^2)^2}\left[\frac{4\pi^2}{l^2}\delta(\bold{k}_\parallel-\bold{k}_{\rm L, \parallel})\delta(\bold{k}^\prime_\parallel-\bold{k}_{\rm L, \parallel}) - \delta(\bold{k}_\parallel - \bold{k}^\prime_\parallel)\right].
\label{sigma dag O}
\end{align}
Notably, Equation~\eqref{sigma dag sigma dag_1} can be written as
$\langle \sigma^\dagger_{\bold{k}_\parallel} \sigma^\dagger_{2\bold{k}^\prime_{\rm L,\parallel} - \bold{k}_\parallel} \rangle_{\rm ss} \propto (\Omega^*)^2/(\Delta - {\rm i}\gamma_0/2)^2$ for small drivings, which is the same result as the one we obtain for a single CE. In contrast, for large drivings, $\langle \sigma^\dagger_{\bold{k}_\parallel} \sigma^\dagger_{2\bold{k}_{\rm L,\parallel} - \bold{k}_\parallel^\prime} \rangle_{\rm ss} \propto 0$, reminiscent of the behaviour of a single QE. The correlator of Equation~\eqref{sigma dag O} also tends to zero for large $\Omega$. Finally, $\bold{v}_{\rm ss}$ can be written as
\begin{equation*}
    \bold{v}_{\rm ss}=\begin{pmatrix}
                \dfrac{32|\Omega|^4}{(\gamma_0^2+4\Delta^2+8|\Omega|^2)^2}\\
                \dfrac{4\left(\Omega^*\right)^2(\gamma_0-{\rm i}2\Delta)^2}{(\gamma_0^2+4\Delta^2+8|\Omega|^2)^2}\\
                \dfrac{-8\Omega^*|\Omega|^2(\gamma_0-{\rm i}2\Delta)}{(\gamma_0^2+4\Delta^2+8|\Omega|^2)^2}
                \end{pmatrix}
\delta(\bold{k}_\parallel-\bold{k}_{\parallel}^\prime),
\end{equation*}
which is the same result as the one we obtain for a single QE, except for the Dirac deltas in $\bold{k}_\parallel$. The power spectrum of the lattice of QEs is finally given by
\begin{equation}
     \mathcal{S}_{\bold{k}_\parallel,\bold{k}_\parallel^\prime}(\omega)=\frac{4\pi^2}{l^2}\frac{4|\Omega|^2(\gamma_0^2+4\Delta^2)}{(\gamma_0^2+4\Delta^2+8|\Omega|^2)^2}\delta(\bold{k}_\parallel-\bold{k}_{\rm L,\parallel})\delta(\bold{k}_\parallel^\prime-\bold{k}_{\rm L, \parallel})\delta(\omega-\omega_{\rm L})+\mathcal{S}^{\rm I}_{\mathrm{QE}}(\omega)\delta(\bold{k}_\parallel-\bold{k}_{\parallel}^\prime),
\label{S array}
\end{equation}
where $\mathcal{S}^{\rm I}_{\mathrm{QE}}(\omega)$ is the incoherent part of the power spectrum of a single coherently driven QE, that is, the second line of Equation~\eqref{S casi}. Note that this incoherent emission term is proportional to delta $\delta(\bold{k}_\parallel - \bold{k}^\prime_\parallel)$ and hence, it is not restricted to $\bold{k}_{\rm L, \parallel}$. Consequently, it describes photon emission with any parallel wavevector, even for an array of noninteracting emitters.

\subsection{Mean-Field or Semiclassical Approximation}
In order to include the interactions between emitters and improve our previous results, we can apply a MFA, where we neglect correlations between different emitters by replacing $\langle \sigma_i^\dagger\sigma_j \rangle$  with $\langle\sigma_i^\dagger\rangle\langle \sigma_j \rangle$ for $i\neq j$. This approach is equivalent to the semiclassical approximation that assumes that the steady-state density matrix of the QE lattice is separable, $\rho=\otimes_i\rho_i$. It is not complicated to see that the MFA is equivalent to transforming the interaction part of the Hamiltonian of Equation~\eqref{hamiltonian space} and the Lindblad dissipators as
\begin{align*}
    &\sum_i \sum_{j\neq i} g_{ij}\sigma_i^\dagger\sigma_j  \xrightarrow{\text{MFA}} \sum_i \sum_{j\neq i} g_{ij}\left(\sigma^\dagger_i\langle \sigma_j \rangle + \langle \sigma^\dagger_i \rangle \sigma_j \right)=\sum_i \left(\sum_{j\neq 0} g_{0j}\langle \sigma_{i+j} \rangle \right)\sigma_i^\dagger + \left(\sum_{j\neq 0} g_{0j}\langle \sigma^\dagger_{i+j} \rangle \right)\sigma_i,\\ \nonumber
    &\sum_{i}\sum_{j\neq i}\frac{\gamma_{ij}}{2}(2\sigma_j\rho\sigma_i^\dagger - \{\sigma_i^\dagger \sigma_j,\rho\})\xrightarrow{\text{MFA}}\sum_i\sum_{j\neq i}\frac{\gamma_{ij}}{2}(2\langle\sigma_j\rangle\rho\sigma^\dagger_i + 2\sigma_j\rho\langle \sigma^\dagger_i\rangle - \{ \langle \sigma^\dagger_i \rangle \sigma_j + \sigma^\dagger_i \langle \sigma_j\rangle, \rho\})=\\
    &=\sum_i \left(\sum_{j\neq 0}(-{\rm i})\frac{\gamma_{0j}}{2}\langle \sigma_{i+j} \rangle \right) {\rm i}[\rho, \sigma_i^\dagger] + \left(\sum_{j\neq 0}{\rm i}\frac{\gamma_{0j}}{2}\langle \sigma^\dagger_{i+j} \rangle \right) {\rm i}[\rho, \sigma_i].
\end{align*}
Then, introducing this into Equation~\eqref{lindblad first}, we get
\begin{equation*}
    \dot{\rho}={\rm i}\frac{1}{\hbar}[ \rho, H_{\rm eff} ]+\sum_{i}\frac{\gamma_0}{2}L_{\sigma_i}(\rho),
\end{equation*}
with the effective Hamiltonian
\begin{align*}
    H_{\rm eff}/\hbar=-\Delta\sum_i\sigma_i^\dagger\sigma_i + \sum_i \left[\left(\Omega_{\rm eff}\right)_i \sigma_i^\dagger+\left(\Omega_{\rm eff}^*\right)_i \sigma_i\right],
\end{align*}
and the effective coherent driving 
\begin{equation}
    \left(\Omega_{\rm eff}\right)_i=\Omega e^{{\rm i}\bold{k}_{\rm L, \parallel}\cdot \bold{r}_i} + \sum_{j\neq 0}\left(g_{0j}-{\rm i}\frac{\gamma_{0j}}{2}\right )\langle \sigma_{i+j} \rangle = \Omega e^{{\rm i}\bold{k}_{\rm L, \parallel}\cdot \bold{r}_i} - \frac{\omega_0^2 \mu^2}{\hbar\varepsilon_0 c^2}\sum_{j\neq0}\hat{\boldsymbol{\mu}}\cdot\bold{G}(|\bold{r}_j|,\omega_0)\cdot \hat{\boldsymbol{\mu}}\langle \sigma_{i+j} \rangle.
\label{effective field}
\end{equation}

The MFA transforms the system of interacting QEs into an effective system of non-interacting ones, with a coherent drive that, importantly, incorporates the effect of the interactions between them. The solution for $\langle \sigma_i \rangle_{\rm ss}$ is, hence,  given by
\begin{equation}
    \langle \sigma_i \rangle_{\rm ss}=\frac{-{\rm i} 2(\gamma_0+{\rm i}2\Delta)\left(\Omega_{\rm eff}\right)_i}{\gamma_0^2 + 4\Delta^2 + 8 |\left(\Omega_{\rm eff}\right)_i|^2}.
\label{sigma i general}
\end{equation}
Equation~\eqref{sigma i general}, together with Equation~\eqref{effective field}, defines an implicit nonlinear relation for $(\Omega_{\rm eff})_i$. In order to solve it, we need to take into account that, since all the QEs are identical and the incident laser is the same for all of them, the effective driving amplitudes can only differ by a phase, i.e.,  $|\left(\Omega_{\rm eff}\right)_i|=|\left(\Omega_{\rm eff}\right)_j|=|\Omega_{\rm eff}|$ for all $i,j$. Substituting $\left(\Omega_{\rm eff}\right)_{i}$ recursively, we obtain the following geometric sum:
\begin{align}\nonumber
    \left(\Omega_{\rm eff}\right)_{i}&=\Omega e^{{\rm i}\bold{k}_{\rm L, \parallel}\cdot \bold{r}_i} \sum_{j=0}^{\infty}\left[\frac{{\rm i}2(\gamma_0+{\rm i}2\Delta)}{\gamma_0^2 + 4\Delta^2 + 8 |\Omega_{\rm eff}|^2}\right]^j \left[\frac{\omega_0 ^2 \mu^2}{\hbar \varepsilon_0 c^2}\mathcal{G}(\bold{k}_{\rm L, \parallel},\omega_0)\right]^j=\frac{\Omega e^{{\rm i}\bold{k}_{\rm L, \parallel}\cdot \bold{r}_i}}{1-\left( \frac{{\rm i}2(\gamma_0+{\rm i}2\Delta)}{\gamma_0^2 + 4\Delta^2 + 8 |\Omega_{\rm eff}|^2} \right)\left(\frac{\omega_0 ^2 \mu^2}{\hbar \varepsilon_0 c^2}\right)\mathcal{G}(\bold{k}_{\rm L, \parallel},\omega_0)}=\\
    &=\Omega_{\rm eff}(\bold{k}_{\rm L, \parallel},\omega_0)e^{{\rm i}\bold{k}_{\rm L, \parallel}\cdot \bold{r}_i}.
\label{field amplitude}
\end{align}

The numerical solution of Equation~\eqref{field amplitude}, consistent with the results in Refs.~\cite{Parmee21_PRA, Ruostekoski23_PRA}, reveals a bistable behavior in $|\Omega_{\rm eff}|$ within a certain range of lattice parameters.
In addition, since $\left(\Omega_{\rm eff}\right)_i=\Omega_{\rm eff}(\bold{k}_{\rm L, \parallel},\omega_0)e^{{\rm i}\bold{k}_{\rm L, \parallel}\cdot \bold{r}_i}$, the solutions for the populations and the power spectrum of the lattice remain unchanged, except that $\Omega$ is replaced by $\Omega_{\rm eff}$.

For completeness, we also present the solution for a lattice of CEs using the MFA. Importantly, in this case, this approximation coincides with the exact result since for CEs, $\langle a^\dagger_i a_j \rangle_{\rm ss} = \langle a^\dagger_i\rangle_{\rm ss}\langle a_j \rangle_{\rm ss}$. Combining Equation~\eqref{effective field} and the solution of $\langle a_i \rangle_{\rm ss}$ for non-interacting CEs 
\begin{equation}
    \langle a_i \rangle_{\rm ss}=\frac{(\Omega_{\rm eff})_i}{\Delta + {\rm i}\gamma_0/2},
\label{a i MFA}
\end{equation}
we obtain the following expression for the effective field:
\begin{equation*}
    \left(\Omega_{\rm eff}\right)_i=\Omega e^{{\rm i}\bold{k}_{\rm L, \parallel}\cdot \bold{r}_i}- \frac{\omega_0^2 \mu^2}{\hbar \varepsilon_0 c^2}\sum_{j\neq0}\hat{\boldsymbol{\mu}}\cdot\bold{G}(|\bold{r}_j|,\omega_0)\cdot \hat{\boldsymbol{\mu}}\frac{\left(\Omega_{\rm eff}\right)_{i+j}}{\Delta + {\rm i}\gamma_0/2}=\frac{\Omega e^{{\rm i}\bold{k}_{\rm L, \parallel}\cdot \bold{r}_i} (\Delta + {\rm i}\gamma_0/2)}{\Delta_{\bold{k}_{\rm L, \parallel}} + {\rm i}\gamma_{\bold{k}_{\rm L, \parallel}}/2}.
\end{equation*}
Therefore, we can rewrite Equation~\eqref{a i MFA} as
\begin{equation*}
    \langle a_i \rangle_{\rm ss}=\frac{\Omega}{\Delta_{\bold{k}_{\rm L, \parallel}} + {\rm i}\gamma_{\bold{k}_{\rm L, \parallel}/2}}e^{{\rm i}\bold{k}_{\rm L, \parallel} \cdot \bold{r}_i}
\end{equation*}
and thus
\begin{equation*}
    \langle a_{\bold{k}_\parallel} \rangle_{\rm ss}=\frac{l}{2\pi}\sum_i \frac{\Omega}{\Delta_{\bold{k}_{\rm L, \parallel}} + {\rm i}\gamma_{\bold{k}_{\rm L, \parallel}/2}}e^{{\rm i}\bold{k}_{\rm L, \parallel} \cdot \bold{r}_i}e^{-{\rm i}\bold{k}_\parallel \cdot \bold{r}_i}=\frac{\Omega}{\Delta_{\bold{k}_{\rm L, \parallel}} + {\rm i}\gamma_{\bold{k}_{\rm L, \parallel}/2}}\frac{2\pi}{l}\delta(\bold{k}_\parallel-\bold{k}_{\rm L, \parallel}),
\end{equation*}
which coincides with Equation~\eqref{a k exact}.

\subsection{Poynting Vector for a Single Quantum Emitter}

Once we have analyzed the populations and power spectra for both the CE and QE lattices, we turn our attention to the particular characteristics of the light radiated by these systems. To that end, we calculate the Poynting vector operator for the lattice. Our starting point is the decomposition of the electric field operator in terms of positive and negative frequency components
\begin{equation*}
   \bm{\mathcal{E}}(\bold{r}, t)=  \bm{\mathcal{E}}^{(+)}(\bold{r}, t) +  \bm{\mathcal{E}}^{(-)}(\bold{r}, t),
\label{E total}
\end{equation*}
Following Ref.~\cite{Glauber63_PR}, the positive and negative frequency components can be expanded as
\begin{equation}
    \bm{\mathcal{E}}^{(+)}(\bold{r}, t)=\int_{0}^\infty {\rm d}\omega \bm{\mathcal{E}}(\bold{r}, \omega) e^{-{\rm i}\omega t},
\label{E +}
\end{equation}
and
\begin{equation*}
    \bm{\mathcal{E}}^{(-)}(\bold{r}, t)=\left [ \bm{\mathcal{E}}^{(+)}(\bold{r}, t)\right]^{\dagger}=\int_{0}^\infty {\rm d}\omega \bm{\mathcal{E}}^\dagger(\bold{r}, \omega) e^{{\rm i}\omega t},
\label{E -}
\end{equation*}
where we have used $\bm{\mathcal{E}}(\bold{r}, -\omega)=\bm{\mathcal{E}}^\dagger(\bold{r}, \omega)$. Furthermore, using the Green tensor in free space \cite{Novotny12_book}, we can write
\begin{equation}
    \bm{\mathcal{E}}(\bold{r}, \omega)=\frac{\omega^2}{\varepsilon_0 c^2}\bold{G}(|\bold{r}-\bold{r}_0|, \omega)\boldsymbol{\mu}\,\sigma(\omega)=\bold{E}(\bold{r},\omega)\sigma(\omega) \text{ for $\omega>0$},
\label{E w}
\end{equation}
where 
\begin{equation}
    \sigma(\omega)=\frac{1}{2\pi}\int_{-\infty}^\infty {\rm d}t \,\sigma(t) e^{{\rm i}\omega t} \text{ for $\omega>0$},
\label{sigma w}
\end{equation}
\begin{equation}
    \sigma^\dagger(\omega)=\frac{1}{2\pi}\int_{-\infty}^\infty {\rm d}t \,\sigma^\dagger(t) e^{-{\rm i}\omega t} \text{ for $\omega>0$}.
\label{sigma dagger w}
\end{equation}
Similarly, for the corresponding magnetic field, we have
\begin{equation}
    \bm{\mathcal{H}}(\bold{r}, \omega)=-{\rm i}\omega \left[\bold{\nabla}\times \bold{G}(|\bold{r}-\bold{r}_0|, \omega)\right]\boldsymbol{\mu}\sigma(\omega)=\bold{H}(\bold{r},\omega)\sigma(\omega) \text{ for $\omega>0$}.
\label{H w}
\end{equation}
In these expressions, $\bold{E}(\bold{r},\omega)$ and $\bold{H}(\bold{r},\omega)$ represent the classical electric and magnetic fields (in frequency domain) created by a dipole moment $\boldsymbol{\mu}$. In addition, it is important to note that the relations $\bm{\mathcal{E}}(\bold{r}, -\omega)=\bm{\mathcal{E}}^\dagger(\bold{r}, \omega)$ and $\bold{E}(\bold{r}, -\omega)=\bold{E}^*(\bold{r}, \omega)$ imply $\sigma(-\omega)=\sigma^\dagger(\omega)$.

We define the Poynting vector operator as $\bold{S}(\bold{r}, t)=\frac{1}{2}\left[\bm{\mathcal{E}}(\bold{r}, t)\times \bm{\mathcal{H}}(\bold{r}, t)+\left(\bm{\mathcal{E}}(\bold{r}, t)\times \bm{\mathcal{H}}(\bold{r}, t)\right)^\dagger \right]$ to ensure it is hermitic. The expected value of this operator is given by
\begin{align*}\nonumber
     \langle \bold{S}(\bold{r}, t) \rangle=\frac{1}{2}\langle \bm{\mathcal{E}}(\bold{r}, t)\times \bm{\mathcal{H}}(\bold{r}, t) \rangle + h.c.={}&\frac{1}{2}\left(\langle \bm{\mathcal{E}}^{(+)}(\bold{r}, t)\times \bm{\mathcal{H}}^{(+)}(\bold{r}, t) \rangle + \langle \bm{\mathcal{E}}^{(+)}(\bold{r}, t)\times \bm{\mathcal{H}}^{(-)}(\bold{r}, t) \rangle \right.\\ \nonumber
    & + \left.  \langle \bm{\mathcal{E}}^{(-)}(\bold{r}, t)\times \bm{\mathcal{H}}^{(+)}(\bold{r}, t) \rangle + \langle \bm{\mathcal{E}}^{(-)}(\bold{r}, t)\times \bm{\mathcal{H}}^{(-)}(\bold{r}, t) \rangle \right) + h.c.
\end{align*}
Next, writing the field operators in frequency domain and making use of Equations~\eqref{E w} and \eqref{H w}, we arrive at
\begin{align}\nonumber
    \langle \bold{S}(\bold{r}, t) \rangle= \left(\frac{1}{2\pi}\right)^2\frac{1}{2}\int_{0}^\infty\int_0^\infty  {\rm d}\omega {\rm d}\omega^\prime & \left[ \bold{E}(\bold{r},\omega)\times \bold{H}(\bold{r},\omega^\prime) \langle \sigma(\omega)\sigma(\omega^\prime)\rangle e^{-{\rm i}\omega t}e^{-{\rm i} \omega^\prime t}\right.\\ \nonumber
    &\left. + \bold{E}(\bold{r},\omega)\times \bold{H}^{*}(\bold{r},\omega^\prime) \langle \sigma(\omega)\sigma^\dagger(\omega^\prime)\rangle e^{-{\rm i}\omega t}e^{{\rm i}\omega^\prime t}\right. \\ \nonumber &+ \left.\bold{E}^{*}(\bold{r},\omega)\times \bold{H}(\bold{r},\omega^\prime) \langle \sigma^\dagger(\omega)\sigma(\omega^\prime)\rangle e^{{\rm i}\omega t}e^{-{\rm i}\omega^\prime t}\right.\\
    & \left. + \bold{E}^{*}(\bold{r},\omega)\times \bold{H}^{*}(\bold{r},\omega^\prime) \langle \sigma^\dagger(\omega)\sigma^\dagger(\omega^\prime)\rangle e^{{\rm i}\omega t}e^{{\rm i}\omega^\prime t}\right] + h.c.
\label{Poynting two ints}
\end{align}
At this point, we need to evaluate the correlators involving $\sigma(\omega)$ and $\sigma^\dagger(\omega)$. Using Equations~\eqref{sigma w} and \eqref{sigma dagger w}, we have
\begin{equation*}
    \langle \sigma^\dagger(\omega)\sigma(\omega^\prime) \rangle=\left(\frac{1}{2\pi}\right)^2\int_{-\infty}^\infty\int_{-\infty}^\infty {\rm d}t{\rm d}\tau \langle \sigma^\dagger(t)\sigma(t+\tau) \rangle e^{{\rm i}(\omega^\prime-\omega)t}e^{{\rm i} \omega^\prime \tau}.
\label{corr sigmas w}
\end{equation*}
In general, the correlator $\langle \sigma^\dagger(t)\sigma(t+\tau) \rangle$ depends on two times, $t$ and $\tau$. However, its steady-state value $ \langle \sigma^\dagger(0)\sigma(\tau) \rangle_{\rm ss}$ only depends on the time difference $\tau$. Therefore, we can write
\begin{align*}
    \langle \sigma^\dagger(\omega)\sigma(\omega^\prime) \rangle_{\rm ss}= \left(\frac{1}{2\pi}\right)^2 \int_{-\infty}^{\infty}e^{{\rm i}(\omega^\prime-\omega)t}{\rm d}t\int_{-\infty}^{\infty} \langle \sigma^\dagger(0)\sigma(\tau) \rangle_{\rm ss} e^{i\omega^\prime \tau}{\rm d}\tau= \mathcal{S}(\omega^\prime)\delta(\omega^\prime-\omega).
\end{align*}
In a similar way, using  $\sigma(-\omega)=\sigma^\dagger(\omega)$, we obtain
\begin{equation*}
    \langle \sigma(\omega)\sigma(\omega^\prime) \rangle_{\rm ss}= \left(\frac{1}{2\pi}\right)^2 \int_{-\infty}^{\infty}e^{{\rm i}(\omega^\prime+\omega)t}{\rm d}t\int_{-\infty}^{\infty} \langle \sigma^\dagger(0)\sigma(\tau) \rangle_{\rm ss} e^{i\omega^\prime \tau}{\rm d}\tau= \mathcal{S}(\omega^\prime)\delta(\omega+\omega^\prime),
\label{sigma sigma}
\end{equation*}
\begin{equation*}
    \langle \sigma(\omega)\sigma^\dagger(\omega^\prime) \rangle_{\rm ss}=\left(\frac{1}{2\pi}\right)^2 \int_{-\infty}^{\infty}e^{{\rm i}(\omega-\omega^\prime)t}{\rm d}t\int_{-\infty}^{\infty} \langle \sigma^\dagger(0)\sigma(\tau) \rangle_{\rm ss} e^{-i\omega^\prime \tau}{\rm d}\tau=\mathcal{S}(-\omega^\prime)\delta(\omega-\omega^\prime),
\end{equation*}
\begin{equation*}
    \langle \sigma^\dagger(\omega)\sigma^\dagger(\omega^\prime) \rangle_{\rm ss}= \left(\frac{1}{2\pi}\right)^2 \int_{-\infty}^{\infty}e^{-{\rm i}(\omega+\omega^\prime)t}{\rm d}t\int_{-\infty}^{\infty} \langle \sigma^\dagger(0)\sigma(\tau) \rangle_{\rm ss} e^{-i\omega^\prime \tau}{\rm d}\tau= \mathcal{S(-\omega^\prime)}\delta(\omega+\omega^\prime).
\label{sigma dag sigma dag}
\end{equation*}
Finally, introducing these correlators into Equation~\eqref{Poynting two ints}, we arrive at
\begin{align*}\nonumber
    \langle \bold{S}(\bold{r}) \rangle_{\rm ss} ={}&\frac{1}{2}\int_{0}^\infty \left[\bold{E}(\bold{r},\omega)\times \bold{H}^{*}(\bold{r},\omega) \mathcal{S(-\omega)} + \bold{E}^{*}(\bold{r},\omega)\times \bold{H}(\bold{r},\omega)\mathcal{S(\omega)} \right]{\rm d}\omega\\ \nonumber
    &+\frac{1}{2}\int_{0}^\infty \left[\bold{E}^{*}(\bold{r},\omega)\times \bold{H}(\bold{r},\omega) \mathcal{S(-\omega)} + \bold{E}(\bold{r},\omega)\times \bold{H}^{*}(\bold{r},\omega)\mathcal{S(\omega)} \right]{\rm d}\omega\\ 
    ={}&\int_{-\infty}^\infty \mathrm{Re}\left(\bold{E}(\bold{r},\omega)\times \bold{H}^{*}(\bold{r},\omega)\right)\mathcal{S(\omega)}{\rm d}\omega.
\label{Poyting finally}
\end{align*}
Note that only the correlators proportional to $\delta(\omega-\omega')$ contribute to the expectation value of the Poynting vector since the integral in Equation~\eqref{Poynting two ints} runs exclusively over positive frequencies. Therefore, for a single frequency, we have
\begin{align*}
    \langle \bold{S}(\bold{r}, \omega) \rangle_{\rm ss}= \mathrm{Re}\left\{\bold{E}(\bold{r},\omega)\times \bold{H}^{*}(\bold{r}, \omega)\right\}\mathcal{S(\omega)}.
\end{align*}

\subsection{Poynting Vector for a Lattice of Quantum Emitters}

In this section, we derive the Poynting vector operator for the lattice of QEs and calculate the far-field intensity radiated by each BS. The electric field operator in frequency domain is given by
\begin{equation}
    \bm{\mathcal{E}}(\bold{r}, \omega)=\frac{\omega^2}{\varepsilon_0 c^2}\sum_i\bold{G}(|\bold{r}-\bold{r}_i|, \omega)\boldsymbol{\mu}\sigma_i(\omega)=\sum_i\bold{E}_i(\bold{r},\omega)\sigma_i(\omega) \text{ for $\omega>0$},
\label{E in i}
\end{equation}
where $\bold{E}_i(\bold{r},\omega)$ is the classical field created by a dipole located at $\bold{r}_i$. We recall that the conditions $\bm{\mathcal{E}}(\bold{r}, -\omega)=\bm{\mathcal{E}}^\dagger(\bold{r}, \omega)$ and  $\bold{E}_i(\bold{r},-\omega)= \bold{E}_i^*(\bold{r},\omega)$ imply $\sigma_i(-\omega)=\sigma_i^\dagger(\omega)$. Using the definition of $\sigma_{\bold{k}_\parallel}$, Equation~\eqref{E in i} becomes
\begin{align*}
     \bm{\mathcal{E}}(\bold{r}, \omega)=\frac{\omega^2}{\varepsilon_0 c^2}\frac{l}{2\pi}\int_{\rm 1BZ}\mathrm{d}\bold{k}_\parallel \boldsymbol{\mathcal{G}}(\bold{r},\bold{k}_\parallel, \omega)\boldsymbol{\mu}\sigma_{\bold{k}_{\parallel}}(\omega)=\frac{l}{2\pi}\int_{\rm 1BZ}\mathrm{d}\bold{k}_\parallel \bold{E}_{\bold{k}_{\parallel}}(\bold{r},\omega)\sigma_{\bold{k}_{\parallel}}(\omega),
\end{align*}
where $\boldsymbol{\mathcal{G}}(\bold{r},\bold{k}_\parallel, \omega)=\sum_i\bold{G}(|\bold{r}+\bold{r}_i|, \omega)e^{{-\rm i}\bold{k}_{\parallel}\cdot\bold{r_i}}$ is the displaced lattice sum and $\bold{E}_{\bold{k}_\parallel}(\bold{r},\omega)=\frac{\omega^2}{\varepsilon_0 c^2}\boldsymbol{\mathcal{G}}(\bold{r},\bold{k}_\parallel, \omega)\boldsymbol{\mu}$. From the definitions of $\sigma_{\bold{k}_\parallel}(\omega)$ and $\bold{E}_{\bold{k}_\parallel}(\bold{r},\omega)$, we infer that $\sigma^\dagger_{\bold{k}_\parallel}(\omega)=\sigma_{-\bold{k}_\parallel}(-\omega)$ and $\bold{E}_{\bold{k}_\parallel}^*(\bold{r},\omega)=\bold{E}_{-\bold{k}_\parallel}(\bold{r},-\omega)$. In the same way, the corresponding magnetic field reads
\begin{equation*}
    \bm{\mathcal{H}}(\bold{r}, \omega)=-{\rm i}\omega\frac{l}{2\pi}\int_{\rm 1BZ}\mathrm{d}\bold{k}_\parallel[\bold{\nabla}\times\boldsymbol{\mathcal{G}}(\bold{r},\bold{k}_\parallel, \omega)]\boldsymbol{\mu}\sigma_{\bold{k}_{\parallel}}(\omega)=\frac{l}{2\pi}\int_{\rm 1BZ}\mathrm{d}\bold{k}_\parallel\bold{H}_{\bold{k}_\parallel}(\bold{r},\omega)\sigma_{\bold{k}_{\parallel}}(\omega).
\end{equation*}
Therefore, the Poynting vector is
\begin{align}\nonumber
    \langle \bold{S}(\bold{r}, t) \rangle={}&\frac{1}{2}\langle \bm{\mathcal{E}}(\bold{r}, t)\times \bm{\mathcal{H}}(\bold{r}, t) \rangle + h.c. \nonumber \\ ={}& \frac{1}{2}\frac{l^2}{4\pi^2}\int_{0}^\infty\int_0^\infty {\rm d}\omega {\rm d}\omega^\prime \int_{\rm 1BZ}\int_{\rm 1BZ}\mathrm{d} \bold{k}_\parallel \mathrm{d} \bold{k}_\parallel^\prime  \nonumber \\ &\times\left[ \bold{E}_{\bold{k}_{\parallel}}(\bold{r},\omega)\times \bold{H}_{\bold{k}_\parallel^\prime}(\bold{r},\omega^\prime) \langle \sigma_{\bold{k}_\parallel}(\omega)\sigma_{\bold{k}_\parallel^\prime}(\omega^\prime)\rangle e^{-{\rm i}\omega t}e^{-{\rm i} \omega^\prime t}\right. \nonumber \\
    &\left. + \bold{E}_{\bold{k}_{\parallel}}(\bold{r},\omega)\times \bold{H}_{\bold{k}_\parallel^\prime}^{*}(\bold{r},\omega^\prime) \langle \sigma_{\bold{k}_\parallel}(\omega)\sigma^\dagger_{\bold{k}_\parallel^\prime}(\omega^\prime)\rangle e^{-{\rm i}\omega t}e^{{\rm i}\omega^\prime t} \right.\nonumber \\ &+ \left. \bold{E}_{\bold{k}_{\parallel}}^{*}(\bold{r},\omega)\times \bold{H}_{\bold{k}_\parallel^\prime}(\bold{r},\omega^\prime) \langle \sigma^\dagger_{\bold{k}_\parallel}(\omega)\sigma_{\bold{k}_\parallel^\prime}(\omega^\prime)\rangle e^{{\rm i}\omega t}e^{-{\rm i}\omega^\prime t}\right. \nonumber \\
    & \left. + \bold{E}_{\bold{k}_{\parallel}}^{*}(\bold{r},\omega)\times \bold{H}_{\bold{k}_\parallel^\prime}^{*}(\bold{r},\omega^\prime) \langle \sigma^\dagger_{\bold{k}_\parallel}(\omega)\sigma^\dagger_{\bold{k}_\parallel^\prime}(\omega^\prime)\rangle e^{{\rm i}\omega t}e^{{\rm i}\omega^\prime t}\right] + h.c.
\label{Poynting total}
\end{align}
To evaluate the correlators, we use $\sigma^\dagger_{\bold{k}_\parallel}(\omega)=\sigma_{-\bold{k}_\parallel}(-\omega)$. By doing so, and moving back to time domain, while replacing the correlators at two times with their steady-state values, we get:
\begin{align*}
    \langle \sigma_{\bold{k}_\parallel}(\omega)\sigma_{\bold{k}_\parallel^\prime}(\omega^\prime)\rangle_{\rm ss} &= \mathcal{S}_{-\bold{k}_\parallel,\bold{k}_\parallel^\prime}(\omega^\prime)\delta(\omega^\prime+\omega), \\
    \langle \sigma_{\bold{k}_\parallel}(\omega)\sigma^\dagger_{\bold{k}_\parallel^\prime}(\omega^\prime)\rangle_{\rm ss}&=\mathcal{S}_{-\bold{k}_\parallel,-\bold{k}_\parallel^\prime}(-\omega^\prime)\delta(\omega^\prime-\omega),  \\
    \langle \sigma^\dagger_{\bold{k}_\parallel}(\omega)\sigma_{\bold{k}_\parallel^\prime}(\omega^\prime)\rangle_{\rm ss}&=\mathcal{S}_{\bold{k}_\parallel,\bold{k}_\parallel^\prime}(\omega^\prime)\delta(\omega^\prime-\omega), \\
    \langle \sigma^\dagger_{\bold{k}_\parallel}(\omega)\sigma^\dagger_{\bold{k}_\parallel^\prime}(\omega^\prime)\rangle_{\rm ss}&=\mathcal{S}_{\bold{k}_\parallel,-\bold{k}_\parallel^\prime}(-\omega^\prime)\delta(\omega^\prime+\omega). 
\end{align*}
Note that, as in the case of the single QE, only the correlators proportional to $\delta(\omega-\omega')$ contribute to the Poynting vector. Therefore, after introducing them into Equation~\eqref{Poynting total}, we get
\begin{align*}\nonumber
    \langle \bold{S}(\bold{r}) \rangle_{\rm ss}={}& \frac{1}{2}\frac{l^2}{4\pi^2}\int_{0}^\infty {\rm d}\omega \int_{\rm 1BZ}\int_{\rm 1BZ}\mathrm{d} \bold{k}_\parallel \mathrm{d} \bold{k}_\parallel^\prime \nonumber \\ &\times \left[\bold{E}_{\bold{k}_{\parallel}}(\bold{r},\omega)\times \bold{H}_{\bold{k}^\prime_\parallel}^{*}(\bold{r},\omega) \mathcal{S}_{-\bold{k}_\parallel,-\bold{k}_\parallel^\prime}(-\omega) + \bold{E}_{\bold{k}_{\parallel}}^{*}(\bold{r},\omega)\times \bold{H}_{\bold{k}^\prime_\parallel}(\bold{r},\omega) \mathcal{S}_{\bold{k}_\parallel,\bold{k}_\parallel^\prime}(\omega) \right] \\ \nonumber
    &+ \frac{1}{2}\frac{l^2}{4\pi^2}\int_{0}^\infty {\rm d}\omega\int_{\rm 1BZ}\int_{\rm 1BZ}\mathrm{d} \bold{k}_\parallel \mathrm{d} \bold{k}_\parallel^\prime \nonumber \\ &\times\left[\bold{E}_{\bold{k}_{\parallel}}^*(\bold{r},\omega)\times \bold{H}_{\bold{k}^\prime_\parallel}(\bold{r},\omega) \mathcal{S}_{-\bold{k}_\parallel,-\bold{k}_\parallel^\prime}(-\omega)+ \bold{E}_{\bold{k}_{\parallel}}(\bold{r},\omega)\times \bold{H}_{\bold{k}^\prime_\parallel}^{*}(\bold{r},\omega) \mathcal{S}_{\bold{k}_\parallel,\bold{k}_\parallel^\prime}(\omega) \right] \\ 
    ={}&\frac{l^2}{4\pi^2}\int_{-\infty}^\infty {\rm d}\omega \int_{\rm 1BZ}\int_{\rm 1BZ}\mathrm{d} \bold{k}_\parallel \mathrm{d} \bold{k}_\parallel^\prime \mathrm{Re}\left(\bold{E}_{\bold{k}_{\parallel}}(\bold{r},\omega)\times \bold{H}^{*}_{\bold{k}^\prime_\parallel}(\bold{r},\omega)\right) \mathcal{S}_{\bold{k}_\parallel,\bold{k}_\parallel^\prime}(\omega),
\end{align*}
and for a single frequency
\begin{equation}
    \langle \bold{S}(\bold{r}, \omega) \rangle_{\rm ss}=\frac{l^2}{4\pi^2}\int_{\rm 1BZ}\int_{\rm 1BZ}\mathrm{d} \bold{k}_\parallel \mathrm{d} \bold{k}_\parallel^\prime \mathrm{Re}\left(\bold{E}_{\bold{k}_{\parallel}}(\bold{r},\omega)\times \bold{H}_{\bold{k}^\prime_\parallel}^{*}(\bold{r},\omega)\right) \mathcal{S}_{\bold{k}_\parallel,\bold{k}_\parallel^\prime}(\omega),
\label{poyntin casi final}
\end{equation}
which is a result very similar to that obtained for a single QE, but with the double integration over the 1BZ. Importantly, this expression is valid independently of the approximation used to calculate the power spectrum since all the information about the quantum state of the system is included in the function $\mathcal{S}_{\bold{k}_\parallel,\bold{k}_\parallel^\prime}(\omega)$. 

Introducing Equation~\eqref{S array} into Equation~\eqref{poyntin casi final}, we get
\begin{align*}\nonumber
    \langle \bold{S}(\bold{r}, \omega) \rangle_{\rm ss} ={}& \frac{4|\Omega_{\rm eff}|^2(\gamma^2+4\Delta^2)}{(\gamma^2+4\Delta^2+8|\Omega_{\rm eff}|^2)^2} \mathrm{Re}\left(\bold{E}_{\bold{k}_{\rm L, \parallel}}(\bold{r},\omega)\times \bold{H}_{\bold{k}_{\rm L, \parallel}}^{*}(\bold{r},\omega)\right)\delta(\omega-\omega_{\rm L})\\
    &+ \frac{l^2}{4\pi^2}\int_{\rm 1BZ}\mathrm{d}\bold{k}_\parallel \mathrm{Re}\left(\bold{E}_{\bold{k_{\parallel}}}(\bold{r},\omega)\times \bold{H}_{\bold{k_{\parallel}}}^{*}(\bold{r},\omega)\right)\mathcal{S}^{\rm I}_{\rm QE, eff}(\omega),
\label{Poynting casi final}
\end{align*}
where we have already replaced $\Omega$ by $\Omega_{\rm eff}$.

Equipped with the expression for the Poynting vector, we proceed to calculate the intensity emitted by the lattice. We start by computing $\bold{E}_{\bold{k_{\parallel}}}(\bold{r},\omega)\times \bold{H}_{\bold{k_{\parallel}}}^{*}(\bold{r},\omega)$. Expressing the Green tensor in free space as \cite{Novotny12_book},
\begin{equation*}
    \bold{G}(|\bold{r}-\bold{r}_i|, \omega)=(\bold{I}+\tfrac{1}{k^2}\bold{\nabla}\bold{\nabla})\frac{e^{{\rm i}k|\bold{r}-\bold{r}_i|}}{4\pi |\bold{r}-\bold{r}_i|},
\end{equation*}
we can write
\begin{equation*}
    \bold{E}_{\bold{k}_\parallel}(\bold{r},\omega)=\frac{\omega^2}{\varepsilon_0 c^2}\sum_i\bold{G}(|\bold{r}-\bold{r}_i|, \omega)\boldsymbol{\mu}e^{{\rm i}\bold{k}_{\parallel}\cdot\bold{r_i}}=\frac{\omega^2}{\varepsilon_0 c^2}(\bold{I}+\tfrac{1}{k^2}\bold{\nabla}\bold{\nabla})\sum_i \frac{e^{{\rm i}k|\bold{r}-\bold{r}_i|}}{4\pi |\bold{r}-\bold{r}_i|}\boldsymbol{\mu}e^{{\rm i}\bold{k}_{\parallel}\cdot\bold{r_i}}.
\end{equation*}
Using the Weyl identity \cite{Novotny12_book} and the periodic nature of the lattice, we get
\begin{align*}
    \sum_i \frac{e^{{\rm i}k|\bold{r}-\bold{r}_i|}}{4\pi |\bold{r}-\bold{r}_i|}e^{i\bold{k}_{\parallel}\cdot\bold{r_i}}=\frac{\rm i}{8\pi^2}\frac{4\pi^2}{l^2}\sum_{\bold{g}}\frac{e^{{\rm i}(\bold{k}_\parallel+\bold{g})\cdot\bold{R}}e^{{\rm i}k_{z,\bold{g}}|z|}}{k_{z,\bold{g}}},
\end{align*}
where $\bold{R}=\bold{r}-z\hat{z}$. Here, $\sum_{\bold{g}}$ represents the sum over the reciprocal lattice vectors and $k_{z,\bold{g}}=\sqrt{k^2-|\bold{k}_{\parallel}+\bold{g}|^2}$. 
Therefore
\begin{align}
   \bold{E}_{\bold{k}_\parallel}(\bold{r},\omega)=\frac{\omega^2}{\varepsilon_0 c^2}\sum_{\bold{g}}\frac{\rm i}{8\pi^2}\frac{4\pi^2}{l^2}\left[\boldsymbol{\mu}-\frac{\bold{k}_{\bold{g}}(\bold{k}_{\bold{g}}\cdot\boldsymbol{\mu})}{k^2}\right]\frac{e^{{\rm i}(\bold{k}_\parallel+\bold{g})\cdot\bold{R}}e^{{\rm i}k_{z,\bold{g}}z}}{k_{z,\bold{g}}}= \sum_{\bold{g}}\bold{E}_{\bold{g}}(\bold{k}_\parallel,\omega),
\label{E g}
\end{align}
where we have assumed that $z>0$ and we have introduced $\bold{k}_{\bold{g}}=(\bold{k}_\parallel+\bold{g},k_{z,\bold{g}})$. In a similar way, since $\bold{H}_{\bold{k}_\parallel}(\bold{r},\omega)=(-{\rm i}\varepsilon_0 c^2/\omega)\left[\bold{\nabla}\times \bold{E}_{\bold{k}_\parallel}(\bold{r},\omega)\right]$, we can write $\bold{H}_{\bold{k}_\parallel}(\bold{r},\omega)=\sum_{\bold{g}}\bold{H}_{\bold{g}}(\bold{k}_\parallel,\omega)=(\varepsilon_0 c^2/\omega)\sum_{\bold{g}}\bold{k}_{\bold{g}} \times \bold{E}_{\bold{g}}(\bold{k}_\parallel,\omega)$. Then,
\begin{equation}
    \bold{E}_{\bold{k_{\parallel}}}(\bold{r},\omega)\times \bold{H}_{\bold{k_{\parallel}}}^{*}(\bold{r},\omega)=\frac{\varepsilon_0 c^2}{\omega}\sum_{\bold{g},\bold{g^\prime}}\left[\left(\bold{E}_{\bold{g}}(\bold{k}_\parallel,\omega)\cdot\bold{E}^*_{\bold{g^\prime}}(\bold{k}_\parallel,\omega)\right)\bold{k}_{\bold{g^\prime}}^* - \left(\bold{E}_{\bold{g}}(\bold{k}_\parallel,\omega)\cdot\bold{k}_{\bold{g^\prime}}^*\right)\bold{E}^*_{\bold{g^\prime}}(\bold{k}_\parallel,\omega)\right].\label{E times H k}
\end{equation}

To calculate the intensity crossing a plane parallel to the lattice, we integrate over the corresponding surface. As a result, only the terms with $\bold{g}=\bold{g^\prime}$  contribute to the sum in Equation~\eqref{E times H k}, yielding:
\begin{equation*}
    \bold{E}_{\bold{k_{\parallel}}}(\bold{r},\omega)\times \bold{H}_{\bold{k_{\parallel}}}^{*}(\bold{r},\omega)=\frac{\varepsilon_0 c^2}{\omega}\sum_{\bold{g}\in \text{rad}}\bold{k}_{\bold{g}}\left | \bold{E}_{\bold{g}}(\bold{k}_{\parallel},\omega)  \right |^2 ,
\label{ExH g}
\end{equation*}
where the $\sum_{\bold{g}\in \text{rad}}$ is restricted to  the reciprocal vectors that make $k_{z,\bold{g}}$ real, thus corresponding to radiative diffraction orders. The intensity crossing a plane parallel to the lattice per unit  of frequency is then given by
\begin{align}\nonumber
    \langle I(\omega) \rangle_{\rm ss}={}&\frac{\varepsilon_0 c^2}{\omega} \frac{4|\Omega_{\rm eff}|^2(\gamma_0^2+4\Delta^2)}{(\gamma_0^2+4\Delta^2+8|\Omega_{\rm eff}|^2)^2}\sum_{\bold{g}\in \text{rad}}k_{z,\bold{g}} \left | \bold{E}_{\bold{g}}(\bold{k}_{\rm L, \parallel},\omega)  \right |^2 \delta(\omega-\omega_{\rm L})\\
    &+ \frac{\varepsilon_0 c^2}{\omega}\frac{l^2}{4\pi^2}\int_{k_{\parallel}\leq k}\mathrm{d}\bold{k}_\parallel k_{z} \left | \bold{E}_{\bold{0}}(\bold{k}_{\parallel},\omega)  \right |^2 \mathcal{S}^{\rm I}_{\rm QE, eff}(\omega),
\label{Pot per Area}
\end{align}
where we have used the equality $\int_{k_{\parallel}\leq k}\mathrm{d}\bold{k}_\parallel k_{z} \left | \bold{E}_{\bold{0}}(\bold{k}_{\parallel},\omega)  \right |^2=\sum_{\bold{g}\in \text{rad}}\int_{\rm 1BZ}\mathrm{d}\bold{k}_\parallel k_{z,\bold{g}} \left | \bold{E}_{\bold{g}}(\bold{k}_{\parallel},\omega)  \right |^2$. We can further simplify the last term by using Equation~\eqref{E g} and performing the integral to obtain 
\begin{align*}
    \frac{l^2}{4\pi^2}\int_{k_{\parallel}\leq k}\mathrm{d}\bold{k}_\parallel k_{z} \left | \bold{E}_{\bold{0}}(\bold{k}_{\parallel},\omega)  \right |^2=\left(\frac{\omega^2}{\varepsilon_0 c^2}\right)^2\frac{k\mu^2}{12\pi l^2}.
\end{align*}
Therefore, Equation~\eqref{Pot per Area} becomes
\begin{equation}
    \langle I(\omega) \rangle_{\rm ss}= \frac{4|\Omega_{\rm eff}|^2(\gamma_0^2+4\Delta^2)}{(\gamma_0^2+4\Delta^2+8|\Omega_{\rm eff}|^2)^2}\sum_{\bold{g}\in \text{rad}}k_{z,\bold{g}} M_{\bold{g}}(\bold{k}_{{\rm L},\parallel},\omega) \delta(\omega-\omega_{\rm L})+ \frac{\omega^4\mu^2}{12\pi\varepsilon_0 c^3}\frac{1}{l^2} \mathcal{S}^{\rm I}_{\rm QE, eff}(\omega).
\label{I final}
\end{equation}
where $M_{\bold{g}}(\bold{k}_\parallel,\omega)=(\varepsilon_0 c^2/\omega)\left | \bold{E}_{\bold{g}}(\bold{k}_\parallel, \omega) \right |^2$. The result expressed by Equation~\eqref{I final} consists of two terms. The first represents the coherent contribution, which includes all diffraction orders and is restricted to $\omega = \omega_{\rm L}$. This contribution vanishes in the limit $\Omega_{\rm eff} \rightarrow \infty$. In contrast, the second term, which corresponds to a incoherent photon emission at any wavevector within the light cone and frequency, does not vanish under strong driving, but instead saturates to a nonzero value, as shown in Figure~\ref{fig2}(b).

Finally, we define the intensity emitted by each BS as $\langle I(\bold{k}_\parallel,\omega) \rangle_{\rm ss}$, such that $\langle I(\omega) \rangle_{\rm ss}= \int_{k_\parallel < k} \mathrm{d}\bold{k}_\parallel \langle I(\bold{k}_\parallel, \omega) \rangle_{\rm ss}$. Therefore,
\begin{align}
   \langle I(\bold{k}_\parallel, \omega) \rangle_{\rm ss} &= \frac{4|\Omega_{\rm eff}|^2(\gamma_0^2+4\Delta^2)}{(\gamma_0^2+4\Delta^2+8|\Omega_{\rm eff}|^2)^2}\sum_{\bold{g}\in \text{rad}}k_{z,\bold{g}} M_{\bold{g}}(\bold{k}_\parallel,\omega) \delta(\omega-\omega_{\rm L})\delta(\bold{k}_\parallel - \bold{k}_{\rm L, \parallel})+ \frac{l^2}{4\pi^2} k_{z}M_{\bold{0}}(\bold{k}_\parallel,\omega) \mathcal{S}^{\rm I}_{\rm QE, eff}(\omega).
\label{eps k}
\end{align}
Again, the first term in this expression represents a coherent contribution restricted to $\bold{k}_\parallel = \bold{k}_{\rm L,\parallel}$ and $\omega=\omega_{\rm L}$. On the contrary, the second term corresponds to the emission involving all the radiative BSs modulated by the incoherent part of the renormalized power spectrum. The evaluation of Equation~\eqref{eps k} is shown in Figure~\ref{fig2}(a).

%

\end{document}